\documentclass[11pt]{article}
\usepackage[numbers]{natbib}

\usepackage[a4paper,margin=18mm,columnsep=7mm]{geometry}
\usepackage[T1]{fontenc}
\usepackage{lmodern}
\usepackage{microtype}
\usepackage{setspace}
\usepackage{authblk}
\usepackage{graphicx} % Required for inserting images
\usepackage{algorithm}
\usepackage[noend]{algpseudocode}
\usepackage{amsmath, amssymb}
\usepackage{bm} % bold maths
\usepackage{setspace, array}
\usepackage{mathtools}
\usepackage{hyperref}
\usepackage{booktabs}
\usepackage{multirow}

\renewcommand{\thefootnote}{\fnsymbol{footnote}}

\newcommand{\vect}[1]{\bm{#1}}
\DeclareMathOperator{\sgn}{\mathrm{sgn}}

\title{A sliding mode observer perspective on chaos synchronisation of turbulent flows}    

\author[1]{Defne E. Ozan}
\author[1,*]{Andrew Wynn}

\affil[1]{Department of Aeronautics, Imperial College London, UK}
\affil[*]{\textit{Corresponding author:} \texttt{a.wynn@imperial.ac.uk}}

\date{}

\begin{document}
\maketitle
\renewcommand{\thefootnote}{\arabic{footnote}}
\begin{abstract}
State estimation of turbulent flows is challenging as infinitesimal prediction errors grow exponentially due to chaos. However, when signals from one chaotic system (the master) are injected into an identical system with different initial conditions (the slave), the slave can synchronise with the master --- a phenomenon observed in turbulent flows. From a control-theoretic perspective, this synchronisation problem can be viewed as observer design. We first show that synchronisation by direct substitution of observed variables into the slave system is equivalent to a sliding mode observer. This perspective enables the analytical verification of stability as well as the design of feedback for robustness. We focus on reduced-order Navier--Stokes models with quadratic non-linearities, a broad class of systems for which stability can be analysed by seeking Lyapunov functions through parameter-dependent linear matrix inequalities solved via semidefinite programming. Extending this framework, we develop a neural network approach combining a neural ordinary differential equation with a decoder to learn data-driven mappings from observation histories to companion forms, where observer design is greatly simplified. Results are demonstrated on the Lorenz system and a nine-dimensional reduced order model of a turbulent shear flow. This work opens opportunities in the rigorous development of observers for turbulent flows.
\\

\end{abstract}

\section{Introduction}
A wide range of applications from climate modelling to feedback flow control and medical imaging requires the identification of the state of turbulent flow from given sparse observations \citep[][]{chevalier_state_2006,kato_data_2015,carrassi_data_2018,wang_state_2021,dirix_synthesis_2022}. The state estimation of turbulent flows is known to be an extremely difficult task, as infinitesimally small prediction errors can diverge exponentially due to the underlying chaotic dynamics. However, it has been remarkably observed that two chaotic trajectories produced by two identical systems with different initial conditions can be made to synchronise by directly substituting some of the state variables from one of these systems, i.e., the master, into the other system, i.e., the slave, \citep{pecora_synchronization_1990}---here, synchronisation refers to the convergence of the state of the slave system to that of the master system and remaining in step with it thereafter. This observation motivates viewing chaos synchronisation as one possible mechanism for reconstructing turbulent flow states from partial observations. 

Viewed from this perspective, the broader problem of turbulent flow state estimation can be organised according to how the observations are incorporated into a dynamical model: The main strategies include continuous data assimilation, in which observations are continuously fed in the model by either directly substituting the state variables \citep[e.g.,][]{yoshida_regeneration_2005, lalescu_synchronization_2013, wang_synchronization_2022} or by introducing a feedback term that nudges the model towards observations \citep[e.g.,][]{azouani_continuous_2014,clark_di_leoni_synchronization_2020}; variational methods, which formulate state recovery as an optimal control problem \citep[e.g.,][]{le_dimet_variational_1986,symon_data_2017,li_small-scale_2020, wang_variational_2025}; and sequential data assimilation methods, which update the state estimate recursively as new observations become available, such as the ensemble Kalman filter \citep[e.g.,][]{evensen_data_2009,suzuki_estimation_2017,fossella_multiscale_2026}. Another stream of machine learning based methods bypass an explicit dynamical model and instead learn mappings directly from observations to the full flow field, either assuming access to high-resolution data and no physics constraints \citep{fukami_super-resolution_2019,erichson_shallow_2020}, or imposing the governing partial differential equations when high-resolution data are unavailable \citep{gao_super-resolution_2021,kelshaw_physics-constrained_2024,page_super-resolution_2025,mo_reconstructing_2025}.

Among these approaches, direct substitution is closely related to chaos synchronisation. \citet{pecora_synchronization_1990} established the foundations of this phenomenon proposing that synchronisation between two chaotic trajectories occurs when the signs of the Lyapunov exponents of the error subsystem are all negative, and showed this in the master-slave configurations of the prototypical Lorenz and R{\"o}ssler systems. The idea of synchronisation has been subsequently investigated in the context of turbulent flows, which has sought to answer whether small-scale dynamics could be reconstructed from large-scale dynamics or coarse-grained observations. Using direct numerical simulations of Navier--Stokes turbulence, \citet{yoshida_regeneration_2005} demonstrated that small eddies can be recovered from large eddies by continuous data assimilation of large-scale Fourier modes with wave number below a critical wave number. \citet{lalescu_synchronization_2013} found that velocity modes below approximately 20 Kolmogorov dissipation lengths are slaved to the larger scales, with the synchronisation rate exponent scaling with the Kolmogorov dissipation time scale. Considering the turbulent channel flow, \citet{wang_synchronization_2022} investigated whether synchronisation is possible when the flow is unknown in a layer within the domain, in which the reconstruction capability was shown to depend on the orientation, location and size of the unknown layer.  

From a control theory perspective, the problem of chaos synchronisation can be viewed as the design of an observer, in which the slave system acts as an estimator driven by partial measurements of the master system \citep{so_observing_1994, nijmeijer_observer_1997}. The synchronisation can then be posed as a condition that the error dynamics of the estimator is asymptotically stable, or in other words the zero error surface is attractive such that the estimator converges to it and remains there. In this paper, we show that synchronisation achieved by direct substitution of the observed state variables into the slave system is equivalent to a sliding mode observer. In a sliding mode observer, the observables' error is kept on a zero surface with the observer dynamics being effectively constrained to a reduced-order manifold \citep{slotine_sliding_1986}. The reduced-order error dynamics are obtained by applying the idea of equivalent control, which assumes that although the observables' error might oscillate, on average it will be on the zero surface. We show that the condition for chaos synchronisation can then be expressed as finding a Lyapunov function for the equivalent error dynamics of the corresponding sliding mode observer. Combined with computational and data-driven tools, this perspective enables us to analyse and design observers in a systematic way. Beyond direct substitution, this approach leads to the design of feedback mechanisms with desirable properties such as faster convergence or more robust disturbance rejection. 

In addition to the exponential sensitivity of chaos, the state estimation of turbulent flows is made challenging by the necessary high dimensionality of accurate numerical simulations. The limitation imposed by the computational cost of numerical simulations has prompted the development of reduced order models with a lower dimension, which can be employed in real-time applications while maintaining accuracy \citep[e.g.,][]{rowley_model_2017}. This approach becomes viable as the long-term dynamics of turbulent flows are expected to lie on a finite dimensional attracting set. In this paper, we specifically focus on reduced-order models (ROMs) of Navier-Stokes equations, which have quadratic non-linearities, representing a large class of problems. For this class of problems, we show that the search of suitable Lyapunov functions can be posed by formulating linear matrix inequalities with parameter dependence. We sample these inequalities along the chaotic attractor and solve them using semidefinite programming (SDP). This approach can effectively prove the stability of observer error dynamics, i.e., the convergence of the prediction error, for any given sensor setup and observer design without having to run expensive simulations. 

State space transformations to companion forms via a non-linear observability matrix can significantly simplify the design of observer gains \citep{drakunov_sliding_1995}. We explore neural network strategies for efficient data-driven mappings from the observation history to these companion forms, where such transformations may be analytically intractable for high-dimensional systems. This strategy is composed of (i) a neural ordinary differential equation (ODE) that models the evolution of the dynamics in the transformed space of observables and their time derivatives, and (ii) a decoder that learns the transformation from these variables to the original space. If the latter mapping exists, then the convergence of an observer designed in this transformed space guarantees convergence in the original space. Whilst the success of the SDP approach utilising quadratic ROMs relies on the existence of a feasible Lyapunov function, the success of the neural network approach relies on the existence of this mapping, i.e., a condition of observability, which is independent of the type of non-linearity, however requires retraining for any new observation map. 

This paper is organised as follows: In \S \ref{sec:sliding_mode_observers}, we make the connection between chaos synchronisation and sliding mode observers, providing the background. In \S \ref{sec:lorenz_example}, we illustrate these points on the Lorenz system, which serves as a prototypical low-dimensional chaotic system, and design an observer utilising semidefinite programming. In \S \ref{sec:general_equiv_err_rom}, we generalise the formulation of the equivalent error dynamics to reduced-order models with quadratic non-linearities. In this way, we extend the application of the sliding mode observer to a nine-dimensional model of turbulent shear flow \citep{moehlis_low-dimensional_2004} in \S \ref{sec:mfe}. In \S \ref{sec:nonlinear_observability}, we introduce the transformations to the companion form for observables. In \S \ref{sec:neural_network}, we propose a neural network strategy to design a fully data-driven sliding mode observer for chaotic systems and apply it to the discussed examples. \S \ref{sec:conclusion} concludes the paper.

\section{Chaos synchronisation and sliding mode observers}\label{sec:sliding_mode_observers}
% DEO: Add connection to nudging, check if need to add citations
\begin{figure}[tb] \label{fig:sm_schematic}
\includegraphics[width=\linewidth]{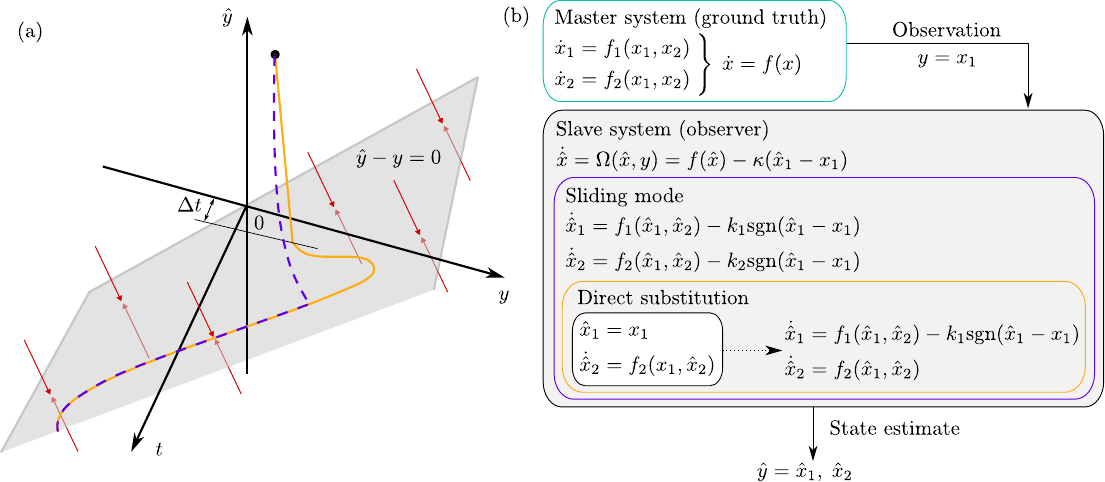}
\caption{(a) Sliding surface, $\hat{y} - y = 0$, for the observed variable $y$ and its estimate $\hat{y}$: Direct substitution (yellow line) reaches sliding in one time step, while sliding mode observer (dashed purple line) ensures that this surface is attracting via feedback of the error sign, $\sgn(\hat{y} - y)$, and choice of feedback gain. (b) Hierarchy of observers: Sliding mode observer generalises direct substitution for chaos synchronisation. }
\end{figure}

Consider a dynamical system $\dot{x} = f(x)$,
\begin{equation}
    f(x) = \begin{pmatrix} f_1(x_1,x_2) \\
  f_2(x_1,x_2) \end{pmatrix} 
    \qquad x = \begin{pmatrix} x_1 \\ x_2 \end{pmatrix}, 
\end{equation}
whose state $x$ is partitioned into `master' modes, $x_1$, and potential `slave' modes $x_2$. The concept of {\em chaos synchronisation} is to assert that if the dynamics of the master modes are given, then the behaviour of the remaining `slave' modes is essentially pre-determined. Mathematically, chaos synchronisation occurs if the state $\hat{x}_2$ of the forced system 
\begin{equation}  \label{eq:synch}
    \dot{\hat{x}}_2 = f_2(x_1, \hat{x}_2),
\end{equation}
obtained by direct substitution of the master modes into the dynamics of the slaves, satisfies 
\begin{equation} \label{eq:synchronisation}
\lim_{t \rightarrow \infty} |\hat{x}_2(t) - x_2(t)|=0,
\end{equation}
from arbitrary initial conditions. There are strong connections between chaos synchronisation and the language of {\em state  observers} for dynamical systems \citep{luenberger_introduction_1971, nijmeijer_observer_1997}, and it is our aim to exploit these connections to gain new insights into both concepts. 

In observer design, the master modes are considered as `observed' or `measured' states, typically denoted $y = x_1$,  which give partial information about the system. An {\em observer} is an auxiliary dynamical system 
\begin{equation} \label{eq:observer}
\dot{\hat{x}} = \Omega(\hat{x},y)
\end{equation}
whose state $\hat{x}$ gives a prediction of the true state, and whose dynamics only depend on the prediction itself and on the observed states $y=x_1$. The aim is to design $\Omega$ such that the convergence property $\lim_{t \rightarrow \infty} |\hat{x}(t)-x(t)|=0$ holds from any initial condition. Typically, this is achieved using dynamics of the form  
\begin{equation} \label{eq:observer_dynamics}
\Omega(\hat{x},y) = f(\hat{x}) - \kappa(\hat{x}_1 - y) 
\end{equation}
which mimic the original system when the observed states agree with their estimations ($y = \hat{x}_1$) but with the feedback $\kappa$ designed\footnote{Implicit in this argument is the assumption that $\kappa(0)=0$.} to ensure convergence in the case of measurable prediction errors. Clearly, a convergent observer has the synchronisation property \eqref{eq:synchronisation} in so far as its unobserved states can be understood purely in-terms of its measured ones. 

There is, however, a subtle distinction between the two concepts. When using an observer of the form \eqref{eq:observer}, a non-zero prediction error $e = \hat{x}-x$ may occur in all states (both observed/masters and unobserved/slaves) at any time, meaning that it is not strictly possible to consider the slave system \eqref{eq:synch} in the form of a full-state observer \eqref{eq:observer}. For {\em sliding mode observers} \citep{slotine_sliding_1986, drakunov_sliding_1995, spurgeon_sliding_2008},  however, this subtle distinction can be removed. A simple implementation is to use the non-linear feedback law 
\begin{equation} \label{eq:sm_ob_feedback}
\kappa(\hat{x}_1 - y) = \begin{pmatrix} k_1 \text{sgn}(\hat{x}_1-y) \\0 \end{pmatrix},
\end{equation}
for some elementwise-positive gain $k_1$. Writing $\Delta f_1= f_1(x_1,x_2) - f_1(\hat{x}_1,\hat{x}_2)$ and $e_1=\hat{x}_1-y$ it follows that the error dynamics for the observed states of the sliding mode observer satisfy 
\begin{equation} \label{eq:sliding_mode_dynamics}
\dot{e}_1 = \Delta f_1 - k_1 \text{sgn}(e_1). 
\end{equation}
If $\Delta f_1$ is bounded---which is often the case for dissipative systems where synchronisation occurs---then for a sufficiently large gain $k_1$ the observer error $e_1$ is driven to zero in finite time\footnote{If $|\Delta f_1|\leq f$ then $|\dot{e}_1| \geq k_1-f >0$ implies a strictly positive error decay rate, bounded away from zero, and this gives finite time convergence.}. After this time, it then follows from \eqref{eq:observer} and \eqref{eq:observer_dynamics} that the observer has dynamics  $\dot{\hat{x}}_2 = f_2(x_1,\hat{x}_2)$ which is precisely the same as the direct substitution system \eqref{eq:synch}.  Consequently chaos synchronisation can be viewed as a special case of non-linear observer design, as indicated schematically in Figure \ref{fig:sm_schematic}.

\subsection{Assisted synchronisation}

The interpretation of chaos synchronisation in the language of observers is of interest for two reasons. First, it is of fundamental interest to be able to classify when synchronisation occurs, and when it does not. Known sufficient conditions for chaos synchronisation involving Lyapunov exponents are hard to check in practice. From the observer perspective, we show that one standard method of verifying observer convergence---by direct construction of Lyapunov functions via semidefinite programming---allows us to practicably verify that chaos synchronisation occurs for a class of quadratic chaotic systems in fluid dynamics. 

A second, and perhaps more wide-reaching, implication is that synchronisation corresponds to only a very particular choice of sliding mode feedback gain, meaning that natural generalisations are possible. Indeed, the simplest generalisation of \eqref{eq:sm_ob_feedback} is to allow error feedback to affect the behaviour of both master and slave states. We shall see that this small additional flexibility allows verification of Lyapunov stability of the corresponding observer error dynamics, even in situations in which pure synchronisation does not hold (or cannot be provably verified) and call this  property {\em assisted synchronisation}. Assisted synchronisation is related to the method of `nudging' from the fluid and weather data assimilation literature. In this method, the model state is nudged toward the observations using a linear feedback term, whose gain serves as a time-relaxation parameter, typically tuned heuristically due to the high dimensionality \citep[e.g.,][]{clark_di_leoni_synchronization_2020}. Here, we exploit the sliding mode feedback structure with switching discontinuity, which is non-linear, to gain insight into the synchronisation dynamics. 

\subsection{Sliding Mode Observers and Non-linear Observability}

Generally, we will consider observer design for systems of the form
\begin{equation} \label{eq:nonlin_sys}
    \dot{x} = f(x), \quad  y = h(x),
\end{equation}
where $x \in \mathbb{R}^n$ is the state vector, $f: \mathbb{R}^{n} \rightarrow \mathbb{R}^n$ is a non-linear function, $y \in \mathbb{R}^p$ is the vector of observable/master states defined by a (possibly non-linear) mapping and $h: \mathbb{R}^{n} \rightarrow \mathbb{R}^p$. For many applications, it is desirable to consider such general observables. However, this is inconvenient from the perspective of analysing observer convergence or chaos synchronisation. 
It is then often preferable to perform a state transformation $z  =H(x)$  so that the first  elements of the transformed state $z$ coincide with the observed/master states. 

When $h$ is linear, this can be achieved with simple linear transformation, as will be detailed subsequently in \S \ref{sec:general_equiv_err_rom}.  In the general non-linear case, finding an appropriate transformation is harder but can still often be performed. To explain, consider the simplest case when $p=1$. Define a state transformation by
\begin{equation} \label{eq:transformation}
z_1 = y, \quad z_2 = \dot{y}, \quad z_3 = \ddot{y}, \quad \cdots \quad z_n = y^{(n-1)},
\end{equation}
under which the system is converted into its `companion form' 
\begin{equation}  \label{eq:companion_form}
    \begin{aligned}
    \dot{z}_i &= z_{i+1}, \quad i = 1, \dots, n-1\\
    \dot{z}_n &= f_n(z), 
    \end{aligned}
\end{equation}
where $f_n$ is a function of the transformed states $z = (z_i)_{i=1}^n$. The existence of $f_n$ in a local neighbourhood of $x$ requires the fulfilment of an observability rank condition \citep{zeitz_observability_1984}. This form is particularly convenient to explain sliding mode design. Letting $e_1 = \hat{x}_1-x_1=\hat{x}_1-y$ be the error in the observed state, we define an observer by
\begin{equation} \label{eq:sliding_observer_companion}
\begin{split}
    \dot{\hat{z}}_i &= \hat{z}_{i+1} - k_i\sgn(e_1), \\
    \dot{\hat{z}}_n &= \hat{f}_n(\hat{z}) - k_n\sgn(e_1),\\
\end{split}
\end{equation}
 and, by the same argument given \S \ref{sec:sliding_mode_observers}, we can achieve zero error in the observed state $y=z_1$ by selecting $k_1$ sufficiently large. Once this convergence is achieved we have $e_1 = \dot{e}_1 = 0$, and the system is said to belong to {\em sliding surface} of the observer.

 The companion form highlights a subtle, but important, advantage of studying synchronisation from the perspective of observer design. On the sliding surface, we are able to explicitly relate the feedback---i.e., the `$\text{sgn}(e_1)$' term---to the unmeasured states of the system. Using \eqref{eq:companion_form} and \eqref{eq:sliding_observer_companion} and applying the notion of `equivalent control', which effectively replaces the discontinuous switching with an average or equivalent continuous signal that achieves the ideal sliding motion, we have $\dot{e}_1 = e_2 - k_1 \text{sgn}(e_1)$ and so, on the sliding surface, it follows that  $\text{sgn}(e_1) = k_1^{-1} e_2$. This relation is particularly valuable since it facilitates an explicit understanding of the influence of the observed/master states on the error dynamics of the unobserved/slave modes, whose dynamics are then largely predetermined through \citep{slotine_sliding_1986,spurgeon_sliding_2008}
\begin{equation}\label{eq:companion_form_error}
    \begin{pmatrix}
        \dot{e}_2 \\ \dot{e}_3 \\\vdots \\ \dot{e}_n 
    \end{pmatrix}
    = 
    \begin{pmatrix}
        -\frac{k_2}{k_1}& 1 & 0 & \cdots & 0 \\
        -\frac{k_3}{k_1}& 0 & 1 & \cdots & 0 \\
        \vdots & 0 & 0 &\ddots & 1 \\
        -\frac{k_n}{k_1}& 0 & 0 & \cdots & 0 \\
    \end{pmatrix}
    \begin{pmatrix}
        e_2 \\ e_3 \\\vdots \\ e_n 
    \end{pmatrix}
    + 
    \begin{pmatrix}
        0 \\ \vdots \\ 0\\  \hat{f}_n(\hat{z})-f_n(z)
    \end{pmatrix}.
\end{equation}
Convergence of the observer error can be analysed in terms of the dependence of the eigenvalues of the system matrix, and the relative size of the gains with respect to $\Delta f_n = f_n(\hat{z})-f_n(z)$. This is often possible to achieve if one can show, or is happy to assume that $\Delta f_n$ is appropriately bounded. 

If convergence is achieved, a crucial point is that this is in the transformed variables $z=H(x)$ only, meaning the real question becomes whether this desired behaviour is inherited by the original system. This is the case if $H$ is invertible, which can be locally checked by the rank condition 
\begin{equation} \label{eq:nl_observability}
\begin{split}
&\text{rank} \left(\frac{\partial H}{\partial x}\right) = n, \\
&H = \begin{pmatrix} h(x) & (L_fh)(x) & \cdots & (L_f^{(n-1)}h)(x) \end{pmatrix}^\top, \quad x \in \mathbb{R}^n,
\end{split}
\end{equation}
where $L_f$ is the Lie derivative operator given by $L_f g:= \nabla f \cdot g$. This rank condition is the standard definition of {\em observability} of the non-linear system \eqref{eq:nonlin_sys} \citep{hermann_nonlinear_1977}. The companion form exists on an open set $S \subset \mathbb{R}^n$ if and only if the map $H$ is injective on $S$, meaning distinct states give distinct vectors of derivatives of the observations, and its Jacobian $\partial H/\partial x$ has rank $n$ at every point of $S$. As a consequence, we have $f_n = (L^{(n)}_fh) \circ H^{-1}$. Here, it is important to stress that we do not require these conditions to hold globally but rather in a subset, which is the chaotic attractor. A second point to emphasise is that the number of derivatives need not be restricted to the state dimension $n$ and in fact, including higher-order information can help distinguishing states.

Consequently, whether (assisted) chaos synchronisation occurs is very closely linked to the non-linear observability of the underlying dynamics with respect to the master modes.  This gives some intuition as to why chaos synchronisation for non-linear systems can occur with relatively few observations. If $f$ is non-linear, then the rank condition \eqref{eq:nl_observability} is far more likely to hold (in comparison with the linear case) due to couplings induced by the Lie derivative operator $L_f$. Indeed, for non-linear $f$, it is believable that such a condition could be true with only a very small number of observed/master modes. 

Unfortunately, due to the process of taking multiple derivatives in the definition of $H$, working with the formal transformation to companion form is not numerically practicable for most applications. In this paper we take two approaches which allow progress. First, we consider a special class of quadratically non-linear systems with linear observation maps, which are of fundamental importance to fluid dynamics. In this case, a simpler state transformation exists to facilitate convergence analysis of sliding mode observers and assisted synchronisation. Second, we show in \S \ref{sec:neural_network} that neural approximations to the companion form transformation are possible, opening the door to enabling sliding mode design on more complex systems.

\section{Checkable conditions for chaos synchronisation using Lyapunov's direct method: a simple example}\label{sec:lorenz_example}

Before discussing the general case, we illustrate the key ideas with a simple example of chaos synchronisation for the Lorenz attractor, a famous 3-state prototypical model for turbulent convection. The dynamics are 
\begin{equation}
    \dot{x} = \underbrace{\begin{pmatrix}
        -\sigma & \sigma & 0 \\
        \rho & -1 & 0 \\
        0 & 0 & -\beta 
    \end{pmatrix}\begin{pmatrix} x_1 \\ x_2 \\ x_3\end{pmatrix} + \begin{pmatrix} 0 \\ -x_1 x_3 \\ x_1 x_2 \end{pmatrix}}_{f(x)}, \qquad x = \begin{pmatrix} x_1 \\ x_2 \\ x_3 \end{pmatrix}, 
    \label{eq:lorenz}
\end{equation}
with parameters chosen as the classical values $\sigma=10$, $\rho=28$, and $\beta=8/3$ which give rise to chaotic dynamics. 
%
% Need to add back somewhere nearer to the results?
%
%We obtain trajectories by integrating the set of ODEs \eqref{eq:lorenz} using a fourth-order Runge-Kutta (RK4) scheme with a time step size of $\Delta t = 0.001$.
%
We assume that the first state $y=x_1$ is observed and ask if synchronisation of the remaining states occurs.

Letting $\hat{x} \in \mathbb{R}^3$ be the observer states and $e = \hat{x}-x$ its error dynamics, the corresponding sliding mode observer satisfies
\begin{equation}
    \begin{split}
    &\dot{\hat{x}} = f(\hat{x}) - \boldsymbol{k}\, \text{sgn}(\hat{x}_1-x_1), \quad \text{with} \\
    &\dot{e} = \begin{pmatrix}
         \sigma(e_2-e_1)  \\
        e_1(\rho-\hat{x}_3)- x_1 e_3 - e_2  \\
        \hat{x}_2 e_1+ x_1 e_2-\beta e_3 
    \end{pmatrix} - \boldsymbol{k} \, \text{sgn}(e_1),
    \end{split}
    \label{eq:lorenz_observer}
\end{equation}
where $\boldsymbol{k}=\begin{pmatrix} k_1 & k_2 & k_3\end{pmatrix}^\top$ are the observer gains. Choosing $k_1$ sufficiently large such that 
\begin{equation}
            \sigma|e_2-e_1| < k_1,
\end{equation}
ensures that $\dot{e}_1 <0$ and the system will be driven to its sliding surface. This requires the mild assumption that $|e_2-e_1|$ is eventually bounded, which can be expected given that the dynamics have an attractor contained within a compact subset of $\mathbb{R}^3$. 

On the sliding surface, where $e_1=\dot{e}_1=0$, we apply the principle of `equivalent control' and infer that  $\sgn(e_1) = \sigma e_2/k_1$. Writing $\tilde{e} = \begin{pmatrix} e_2 & e_3 \end{pmatrix}^\top$, $\boldsymbol{s} = \begin{pmatrix} \sigma/k_1 & 0 \end{pmatrix}$, and  $\tilde{\boldsymbol{k}} = \begin{pmatrix} k_2 & k_3 \end{pmatrix}^\top$, we can express the error dynamics of the unobserved modes as 
%\begin{align}\label{eq:lorenz_equivalent_error}
%    \dot{e}_y &= -\left(1+\frac{\sigma k_y}{k_x}\right)e_y - x e_z , \\
%    \dot{e}_z &= \left(x-\frac{\sigma k_z}{k_x}\right)e_y -\beta e_z 
%\end{align}
\begin{equation} \label{eq:lorenz_equivalent_error}
    \dot{\tilde{e}} = (A(x_1)- \tilde{\boldsymbol{k}} \boldsymbol{s})\tilde{e}, \quad \text{where} \quad A(x_1) =
    \begin{pmatrix}
        -1 & -x_1 \\
         x_1 & -\beta 
    \end{pmatrix},
\end{equation}
Studying pure synchronisation, i.e. when $k_2=k_3=0$, is now simple. We have 
\[
\frac12 \frac{d}{dt} \| \tilde{e}\|_2^2  = - e_2^2 - \beta e_3^2 \leq -\|\tilde{e}\|_2^2,
\]
whence $\lim_{t \rightarrow \infty} \|\tilde{e}\|_2 = 0$, which means that synchronisation provably occurs. 

In the language of non-linear systems theory,  $V(\tilde{e}) = \frac12 \|\tilde{e}\|^2$ is called a {\em Lyapunov function} for the error dynamics on the sliding surface in the sense that it is strictly positive definite, radially unbounded, and satisfies the decay property $\dot{V} < 0$ along trajectories of the system. In general, for any globally stable non-linear system there is a Lyapunov function satisfying these properties which provides a {\em certificate of convergence} for the system. This gives, at least theoretically, a constructive way to check if chaos synchronisation occurs. Unfortunately, there is no reason to expect that  Lyapunov functions will always be as simple the one  used above. In general, both constructing Lyapunov functions or, alternatively, having confidence that no such function exists presents a serious practical challenge. 

A well-known path forward is to use a simple, yet powerful, generalisation to general quadratic Lyapunov functions. The idea is to consider $V(\tilde{e}) = \tilde{e}^\top P \tilde{e}$, parametrised by a symmetric, strictly positive definite matrix $P \in \mathbb{S}_{++}^n$. Such functions are automatically positive definite, radially unbounded and, importantly, the decay condition $\dot{V} <0$ is often tractable to check computationally. For the Lorenz error dynamics \eqref{eq:lorenz_equivalent_error}, for example, the decay condition holds if\footnote{The notation $Q \prec 0$ means strictly negative definite, i.e. $x^\top Q x < 0$ for any $x \neq 0$.}
\begin{equation} \label{eq:lorz_lyap}
    P(A(x_1)-\tilde{\boldsymbol{k}} \boldsymbol{s}) + (A(x_1)-\tilde{\boldsymbol{k}} \boldsymbol{s})^\top P \prec 0,
\end{equation}
for any $x_1$. This equation has three very useful properties. 

First, if \eqref{eq:lorz_lyap} holds at two values of $x_1 = s_-$ and $x_1 = s_+$ then, since $A(\cdot)$ is affine, it must also hold for any $s_- \leq x_1 \leq s_+$. Since we know that $x_1$ is eventually bounded for any solution to the Lorenz dynamics (it is observed that $-16 \leq x_1 \leq 16$, for example) the decay condition $\dot{V}<0$ can be checked by verifying only {\em two} constraints of the form \eqref{eq:lorz_lyap}. Second, the condition \eqref{eq:lorz_lyap} is linear in $P$ with such an equation typically called a Linear Matrix Inequality (LMI). As a consequence, searching for a matrix $P$ to satisfy a finite number of LMI constraints is a convex optimization problem, and falls under the general class of problems known as Semidefinite Programmes (SDPs). These can be solved efficiently with existing compuational tools for moderate sized problems. Third, after making the transformation $\boldsymbol{r} = P\tilde{\boldsymbol{k}}$, the LMI \eqref{eq:lorz_lyap} becomes linear in both the Lyapunov function parameters $P$ and in the observer gain variables. This means that the search for an {\em assisted synchronisation} strategy can also be systematically implemented using SDPs. 

To complete this example, we solve the SDP associated with searching for $P$ and $\tilde{\boldsymbol{k}}$ to satisfy \eqref{eq:lorz_lyap} with $x_1 = \pm 16$. In fact, we go further and replace the right-hand-size of \eqref{eq:lorz_lyap} by $-2\gamma P$ to enforce the stronger exponential error decay condition $\dot{V} \leq -2 \gamma V$ \citep{boyd_linear_1994}. The search can then be performed to maximize $\gamma$. We implement this SDP in YALMIP \citep{lofberg_yalmip_2004} on MATLAB using the MOSEK 11.2 \citep{mosek_aps_mosek_2026} solver to obtain gains $\tilde{k}_2 = 7.996$ and $\tilde{k}_3 = 1.595$.

\begin{figure}[tb]
    \includegraphics[width=\linewidth]{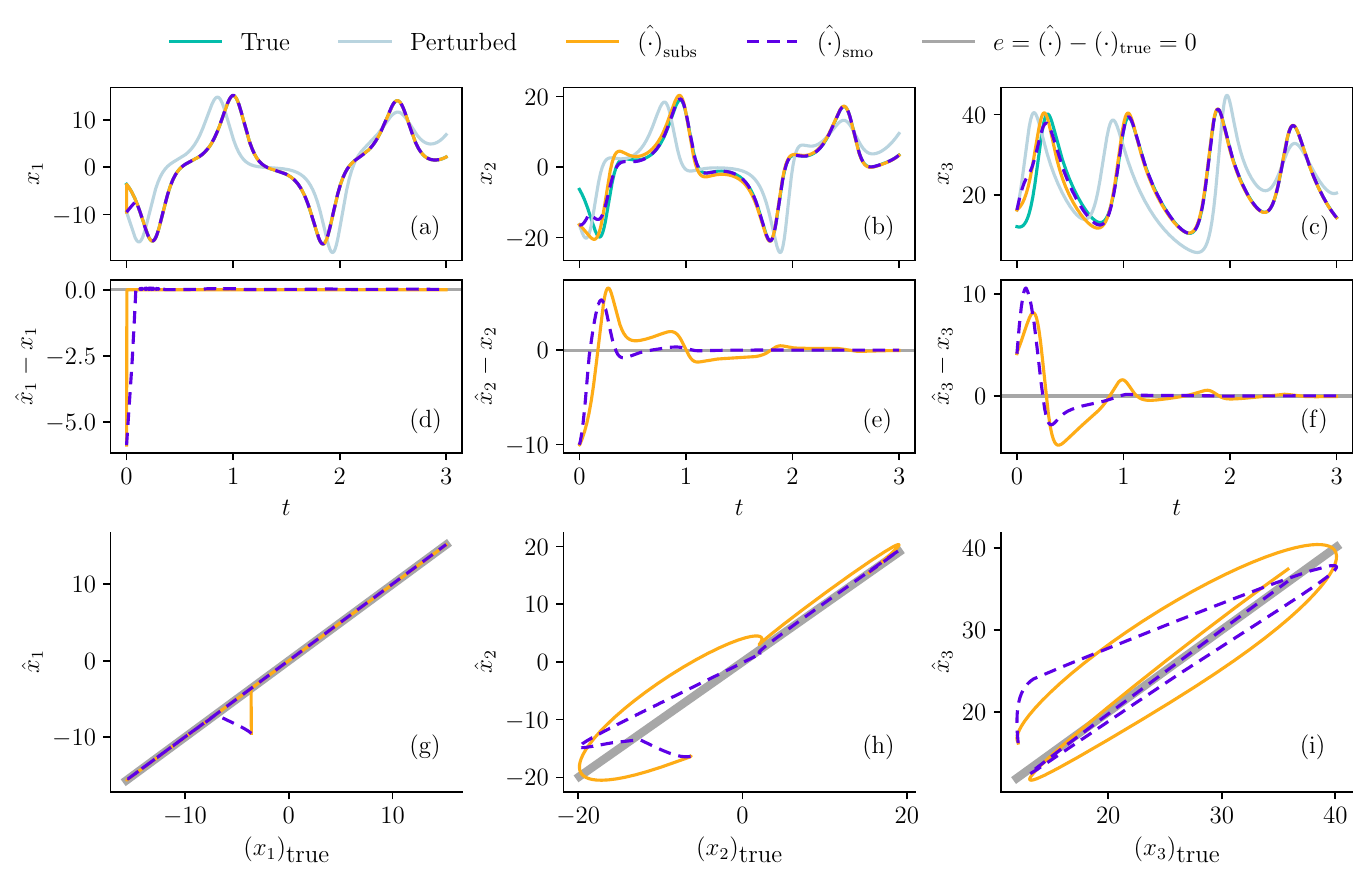}
    \caption{State estimation of the Lorenz system from $x_1$ observations using the direct substitution method, $\hat{(\cdot)}_{\mathrm{subs}}$, and the sliding mode observer designed with the semidefinite programming approach \eqref{eq:lorz_lyap}, $\hat{(\cdot)}_{\mathrm{smo}}$. Shown (a -- c) the state variables, (d -- e) the estimation error, and (g -- h) phase plots of the estimated variables $\hat{(\cdot)}$ vs the ground truth $(\cdot)_{\text{true}}$. Observer design enables quicker convergence.}\label{fig:lorenz_pred_x}
\end{figure}

 Trajectories of the corresponding sliding mode observer, initialised from a random initial condition, are shown in Figure \ref{fig:lorenz_pred_x} for both the direct substitution method, $\hat{(\cdot)}_{\mathrm{subs}}$, and the designed sliding mode observer with non-zero gains, $\hat{(\cdot)}_{\mathrm{smo}}$. To numerically implement the sliding mode observer, we replace the sign operation with a saturation function 
\begin{equation}\label{eq:saturation}
    \mathrm{sat}(e) = \left\{ \begin{array}{rcl}
        \sgn(e) && \text{if} \quad  |e| \geq \epsilon, \\
        e/\epsilon &&  \text{if} \quad |e| < \epsilon,
    \end{array} \right.
\end{equation}
and set $ \epsilon = 10^{-2}$, which eliminates the chattering behaviour. 

During the integration with the fourth-order Runge-Kutta (RK4) scheme, we feed the observations in the observer in a first-order hold fashion, i.e., as a linear interpolation between the time steps $t_k$ and $t_{k+1}$. For the sliding mode observer, we choose $k_1 = 100$, which forces the observer onto the sliding surface after approximately 0.1 time units for the considered initial condition, in contrast to the substitution based approach in which $e_1=0$ is enforced after a single step. Interestingly, while both approaches achieve synchronisation, convergence is slower in the `slave' states $x_2,x_3$ for the direct substitution approach, as can be seen in Figure \ref{fig:lorenz_pred_x}~(e-f), highlighting the potential advantage of assisted synchronisation, which has imposed the stronger exponential decay condition. When only $x_2$ is observed, synchronisation also occurs and can further be proven through the described SDP procedure. In contrast, for most initial conditions that we have tested, observing only $x_3$ did not lead to synchronisation; and consistent with this result, the SDP could not find a feasible solution either.

%Subsequently, the other state variables also converge to their true values with a faster rate compared to the direct substitution, which in contrast displays oscillations around $e_y = e_z = 0$. The gain $k_x$ can be adjusted to achieve sliding as quickly as needed depending on the initial error in $x$. While a too low $k_x$ will not satisfy the sliding condition, a too high $k_x$ may cause chattering and instability. While both direct substitution and sliding mode observer achieve synchronisation, with latter, we have been able to improve the convergence. 

\section{Chaos Synchronisation of Dissipative Quadratic Systems with Linear  Observables}\label{sec:general_equiv_err_rom}

We now seek to extend the observations for the Lorenz system to more general quadratically non-linear systems of the form 
\begin{equation}
\begin{aligned}
    \dot{x} &= Lx + N(x)x, \\
    \label{eq:quadratic_rom}
    y &= Cx,
\end{aligned}
\end{equation}
where $x \in \mathbb{R}^{n}$, $L \in \mathbb{R}^{n \times n}$ and the $p$ observable states are given by $C \in \mathbb{R}^{p \times n}$. The non-linear component of the dynamics is represented by symmetric matrices $Q^{(i)} \in \mathbb{S}^n$ such that 
\[
N(x) = \begin{pmatrix} x^\top Q^{(1)} \\ \vdots \\ x^\top Q^{(n)} \end{pmatrix} \in \mathbb{R}^n, \qquad x \in \mathbb{R}^n, 
\]
is a linear operator with the properties that
\begin{equation} \label{eq:N_properties}
N(a)b=N(b)a, \; \text{and} \; N(a)a-N(b)b = N(a+b)(a-b)
\end{equation}
for any $a,b \in \mathbb{R}^n$. Any quadratically non-linear system can be represented in this form. 

We consider a sliding mode observer defined by 
\begin{equation}
    \dot{\hat{x}} = L\hat{x} + N(\hat{x})\hat{x}-K\sgn(C(\hat{x}-x)),
\end{equation}
where $K \in \mathbb{R}^{n \times p}$ is the feedback gain. Letting $e = \hat{x}-x$ be the observer error, and using the properties \eqref{eq:N_properties} of the chosen representation of the non-linear term, we can write the error dynamics as 
\begin{equation} \label{eq:gen_case_error_dyn}
    \dot{e} = Le + N(\hat{x}+ x)e-K\sgn(Ce).
\end{equation}
Assuming that $K$ is designed so that the system is driven to the sliding surface $C\dot{e} = 0$, we obtain the `equivalent control' relation $\text{sgn}(Ce) = (CK)^{-1} C (L + N(\hat{x}+x))e$. Consequently, on the sliding surface the equivalent error dynamics are (see e.g., \citep{slotine_sliding_1986}), 
\begin{equation}
    \dot{e} = (I - K(CK)^{-1} C) (L +  N(\hat{x} + x))e.
\end{equation}

\subsection{A convenient transformation} \label{sec:transformation}

To facilitate the design of $K$ and analysis of the error dynamics, it is convenient to transform the system to explicitly separate the observed and unobserved states. To do this, we propose using the invertible transformation 
\begin{equation} \label{eq:transform}
   T = \begin{pmatrix}
        C \\ \tilde{C}
    \end{pmatrix}, \quad \text{where} \quad  \mathrm{ker}(C) = \mathrm{Im}(\tilde{C}^\top),
\end{equation}
and $\mathrm{ker}(C), \mathrm{Im}(C)$ are the kernel and image of the observation matrix $C$, respectively. The transformed state 
\begin{equation}\label{eq:state_transform}
z = Tx = \begin{pmatrix} y \\ \tilde{C}x \end{pmatrix} 
\end{equation}
is then such that its first $p$ elements coincide with the observed states. The transformed dynamics satisfy 
\begin{equation}
    \dot{z} = \tilde{L}z + \tilde{N}(z)z,
\end{equation}
with $\tilde{L} = TLT^{-1}$ and $\tilde{N}(z) = TN(T^{-1} z)T^{-1}$, while the dynamics of the error $\xi = \hat{z}-z = Te$ on the sliding surface are
\begin{equation}\label{eq:equiv_err_quad}
    \begin{split}
    \dot{\xi} &= T(I - K(CK)^{-1} C) (L +  N(T^{-1}(\hat{z}+z)))T^{-1}\xi, \\
            %&= \matr{T}
            %(\matr{I} - \matr{K}(\matr{C}\matr{K})^{-1} \matr{C}) \matr{T}^{-1} \matr{T} (\matr{L} +  \matr{N}(\matr{T}^{-1}(\hat{\vect{z}}+\vect{z})))\matr{T}
            %^{-1}\vect{\xi}, \\
            &= (I - TK(CK)^{-1} C T^{-1}) (\tilde{L} +  \tilde{N}(\hat{z}+z))\xi, \\
            %&= \left(\matr{I} - \begin{bmatrix}
            %\matr{I}_p \\ \tilde{\matr{C}}\matr{K}(\matr{C}\matr{K})^{-1}
            %\end{bmatrix}
            %\begin{bmatrix}
            %    \matr{I}_p & \matr{0}_{p \times n-p}
            %\end{bmatrix}\right) (\tilde{\matr{L}} +  \tilde{\matr{N}}(\hat{\vect{z}}+\vect{z}))\vect{s}, \\
            & = \begin{bmatrix}
                0_{p \times p} & 0_{p \times n-p} \\
                -\tilde{C}K(CK)^{-1} & I_{n-p}
            \end{bmatrix}(\tilde{L} +  \tilde{N}(\hat{z}+z))\xi,
           % &  = \begin{bmatrix}
           %     0_{p \times p} & 0_{p \times n-p} \\
           %     A_{21} -\tilde{C}K(CK)^{-1} A_{11} & A
           %     _{22} -\tilde{C}K(CK)^{-1} A_{12}
           % \end{bmatrix} \xi,
        \end{split}
\end{equation}
The last line follows from the definition of $T$ \eqref{eq:transform} and the fact that $CT^{-1} = \begin{pmatrix} I_{p} & 0_{n-p} \end{pmatrix}$. 

This makes clear the fact that the error the first $p$ states of $\xi$ (i.e. the observed states $y=Ce$) are zero on the sliding surface. More usefully, we also obtain the dynamics error $\tilde{\xi} = \tilde{C}e \in \mathbb{R}^{n-p}$ for the unobserved states. Writing 
\begin{equation*}
     \tilde{L} + \tilde{N}(\hat{z} +z)  = \begin{pmatrix}
        A_{11}(\hat{z}+z) & A_{12}(\hat{z}+z) \\ A_{21}(\hat{z}+z) & A_{22}(\hat{z}+z)    \end{pmatrix},
        \end{equation*}
where $A_{ij}$ have affine dependence on $z + \hat{z}$, it follows from \eqref{eq:equiv_err_quad} that 
\begin{equation}
    \dot{\tilde{\xi}} = (A_{22}-\underbrace{\tilde{C}K}_{K_a}(\underbrace{CK}_{K_s})^{-1} A_{12})\tilde{\xi}.
    \label{eq:equiv_err_s_tilde}
\end{equation}

It is worth commenting on the different roles of the feedback terms $CK$ and $\tilde{C}K$ in reduced error dynamics \eqref{eq:equiv_err_quad}. The former term, $K_s = CK \in \mathbb{R}^{p \times p}$, is the component of the available gains used to achieve the `sliding' behaviour and perfect synchronisation of the observed/master modes in \eqref{eq:gen_case_error_dyn}.  Once designed, the matrix $K_s$ is assumed to be fixed and this uses $p^2$ of the available degrees of freedom in the feedback term $K \in \mathbb{R}^{n \times p}$. The remaining $p(n-p)$ degrees of freedom can be used to design the `assisted synchronisation' gain  $K_a = \tilde{C}K \in \mathbb{R}^{(n-p) \times p}$, which essentially becomes a free design variable that can be chosen to help stabilize the equivalent error dynamics \eqref{eq:equiv_err_s_tilde}.  

\subsection{Pure synchronisation}

Pure synchronisation corresponds to the choice $K_a=0$, upon which the equivalent error dynamics become 
\begin{equation} \label{eq:red_error_pure}
\tilde{\xi} = A_{22}(\hat{z} + z) \tilde{\xi}.
\end{equation}
As for the Lorenz example discussed in \S \ref{sec:lorenz_example}, sufficient conditions for stability of this parameter-varying system can be found by considering Lyapunov function candidates of the form $V(\tilde{\xi}) = \tilde{\xi}^\top P \tilde{\xi}$ for parameters $P \in \mathbb{S}^{n}_{++}$. In particular, if 
\begin{equation} \label{eq:LMI_pure}
PA_{22}(\hat{z}+z) + A_{22}(\hat{z}+z)^\top P \prec 0
\end{equation}
for any possible $\hat{z}+z$, it follows that $\dot{V} <0$ along trajectories of \eqref{eq:red_error_pure} and that pure synchronisation $\lim_{t\rightarrow \infty} \tilde{\xi} =0$ occurs. 

Verifying the condition \eqref{eq:LMI_pure} is challenging (but not impossible) due to its dependence on both the state and observer dynamics. 
In simple cases, such as for the Lorenz dynamics in \S \ref{sec:lorenz_example}, it can be shown to hold globally. Indeed, we showed that
\begin{equation} \label{eq:A_22_lorenz}
A_{22}(z + \hat{z}) = \begin{pmatrix} -1 & 0 \\ 0 & -\beta \end{pmatrix} + z_1 \begin{pmatrix} 0 & -1 \\ 1 & 0 \end{pmatrix}.
\end{equation}
Since the matrix multiplying $z_1$ is anti-symmetric, it follows that \eqref{eq:LMI_pure} holds for any $z+\hat{z}$ with the choice $P=I$. Hence, pure synchronisation holds globally. 

In general, such neat cancellation is unlikely to occur. To make progress, we focus on applications to the wide class of `dissipative' dynamical systems which, like the Lorenz system, includes many important examples from fluid mechanics. Dissipative systems have a compact {\em absorbing set} $S \subset \mathbb{R}^n$ with the property that, for any initial condition of the system, the state satisfies $z(t) \in S$ for sufficiently large times. Usually, one can also prove that the observer state $\hat{z}(t)$ also has a similar absorbing set. 

This information is useful, given the fact that \eqref{eq:LMI_pure} has {\em affine} dependence on the unknown state $z + \hat{z}$. If we are able to check  that $PA_{22}(s_i) + A_{22}(s_i)^\top P \prec 0$ for a finite number of states $s_i \in \mathbb{R}^n$---which we refer to subsequently as `check points'---it follows that $PA_{22}(s) + A_{22}(s)^\top P \prec 0$ also holds for any convex combination 
\[
s = \sum_{i=1}^N \lambda_i s_i, 
\]
of the check points, i.e., whenever $\sum_i \lambda_i =1$ and $\lambda_i >0$.  Consequently, if one can solve the feasibility problem
\begin{equation} \label{eq:pure_SDP}
    \begin{aligned}
        \text{find} \quad & P \\
        \text{subject to } \quad & PA_{22}(s_i) + A_{22}(s_i)^\top P \prec 0, \quad i=1,\dots,N, \\
        & P \in \mathbb{S}_{++}^n 
    \end{aligned}
\end{equation}
then \eqref{eq:LMI_pure} must hold whenever $z+\hat{z}$ belongs to the {\em convex hull} of the check points $(s_i)_{i=1}^N$. 

Pure synchronisation can then be proven if a set of check points can be found such that \eqref{eq:pure_SDP} is feasible at each point and their convex hull contains the system's absorbing sets. Even in this more general setting, the optimization problem \eqref{eq:pure_SDP} is still a convex  SDP, due to the fact that the constraints are matrix definiteness conditions which have affine dependence on the search variable $P$.

\subsection{Assisted synchronisation}

If pure synchronisation cannot be checked, one may try to make use of the additional `assisted' feedback gain $K_a$. Given a set of check points $(s_i)_{i=1}^N$, and a pre-selected sliding feedback $K_s$, we can search for assisted gains to stabilize the effective error dynamics \eqref{eq:equiv_err_s_tilde} by solving the following SDP:
\begin{equation}\label{eq:pure_SDP_R}
\begin{aligned}
    \text{find}\quad & P,\,R \\
    \text{subject to}\quad
    & PA_{22}(s_i)+A_{22}(s_i)^\top P
      -RK_s^{-1}A_{12}(s_i) \\
    &\quad +A_{12}\left(s_i\right)^\top\left(K_s^\top\right)^{-1}R^\top
      \prec 0,\quad i=1,\ldots,N, \\
    & P\in\mathbb{S}_{++}^{n},\qquad
      R\in\mathbb{R}^{(n-p)\times p}.
\end{aligned}
\end{equation}

Suppose that feasible solutions $P,R$ to this problem can be found. Letting $V(\tilde{\xi}) = \tilde{\xi}^\top P \tilde{\xi}$ and defining the assisted gain $K_a = P^{-1}R$, we have that, along trajectories of \eqref{eq:equiv_err_s_tilde}, 
\[
P\left(A_{22} - K_a K_s^{-1}A_{12}\right) + \left(A_{22} - K_a K_s^{-1} A_{12}\right)^\top P \prec 0
\]
implying $\dot{V} < 0$ whenever $z +\hat{z}$ belongs to the convex hull of the check points $(s_i)_{i=1}^N$. Assisted synchronisation can therefore also be checked by solving a convex SDP and, moreover, it is more likely to be verifiable due to the additional search variable $R$.

\subsection{Choice of `check' points}

For low-dimensional systems it is possible to select a small number of `check points' whose convex hull contains any compact subset of $\mathbb{R}^n$, and thus any likely absorbing set of a dissipative system. For the Lorenz dynamics, for example, analysis of $A_{22}$ in \eqref{eq:A_22_lorenz} only requires knowledge of one state $z_1$. If this is known (or assumed) to belong to an interval of $\mathbb{R}$, then any possible behaviour can be captured using two `check points' chosen at each end of this interval. 

For higher $n$-dimensional systems, however, such an approach would lead to the need for $2^n$ constraints, and even convex SDPs become computationally infeasible at even moderate values of $n$.  In practice, therefore, we propose a more pragmatic approach and choose `check points' sampled from the chaotic attractor of the system. For ergodic systems, it is likely that the convex hull of a moderate number of check points lies very close to a large majority of points on the chaotic attractor. Thus, the LMI constraints in \eqref{eq:pure_SDP}, \eqref{eq:pure_SDP_R} may well provide a close enough approximation to enable the practical design of an observer with convergent error dynamics. 

\section{Chaos synchronisation for a low-dimensional turbulent shear flow model}\label{sec:mfe}

We apply our approach to study chaos synchronisation in a nine-dimensional model of a turbulent shear flow, proposed by \citet{moehlis_low-dimensional_2004}, which  qualitatively captures behaviour associated with turbulent flows such as relaminarisation and turbulent bursts, and has been widely used to study the transition to turbulence \citep{skufca_edge_2006}, predictability of chaos \citep{srinivasan_predictions_2019} and extreme events \citep{racca_data-driven_2022}.

The model describes flow through a channel between two walls, separated by a distance $L_y$. The velocity field $\boldsymbol{u}$ satisfies the incompressible Navier-Stokes equations 
\begin{equation} \label{eq:NSE}
\begin{split} 
\frac{\partial \boldsymbol{u}}{\partial t} + (\boldsymbol{u} \cdot \nabla) \boldsymbol{u} + \nabla p &= \frac{1}{\text{Re}} \nabla^2 \boldsymbol{u} + \boldsymbol{F}(y), \\
\nabla \cdot \boldsymbol{u} & = 0,
\end{split}
\end{equation}
where $\boldsymbol{F}$ is a body force in the streamwise direction $x$---which depends sinusoidally on the wall normal co-ordinate $y$---and serves to drive the flow. Variables are nondimensionalised using $L_y/2$ as a length-scale, and a characteristic velocity scale $U_0$ arising from a steady solution (equilibrium point) of \eqref{eq:NSE}. Free-slip boundary conditions are assumed at the walls, and periodic boundary conditions imposed in the remaining directions. The single governing parameter is the Reynolds number $\text{Re} = U_0 L_y / 2\nu$, which depends on the fluid viscosity $\nu$. Small Reynolds numbers typically give rise to instability and chaos. 

A nine-state model of the flow is given in \citep{moehlis_low-dimensional_2004}, derived by Galerkin projection of the Navier-Stokes equations onto a judiciously-chosen set of Fourier modes. Since \eqref{eq:NSE} is quadratically non-linear, this gives a finite-dimensional system of the form 
\begin{equation} \label{eq:9_state}
    \dot{a} = c + L a + N(a)a,
\end{equation}
with state $a = (a_1,\dots,a_9)^\top$, and constant forcing $c \in \mathbb{R}^9$. The exact form of these equations are specified in the Appendix \ref{sec:appendix}, but we note here that these have the same form as was analysed in \S \ref{sec:general_equiv_err_rom} but with an additional constant forcing. However, in the derivation of the equivalent error, this terms plays no role. It is known (see \citep{wynn_observer_2011}) that the dynamics form a compact absorbing set and we therefore assume without loss of generality that all states are uniformly bounded for all times. 

For the remainder of the paper, we fix the parameters of the system to $L_x = 4 \pi$, $L_z = 2 \pi$ and $Re = 400$, and generate trajectories similarly to \citep{srinivasan_predictions_2019,racca_data-driven_2022}. The set of ODEs \eqref{eq:9_state} (but see also  \eqref{eq:mfe} in the Appendix \ref{sec:appendix}) are integrated with a fourth-order Runge-Kutta scheme using a time step size of $\Delta t = 0.25$. We generate a base trajectory of length 4,200 time units using the initial condition $a_0 = \hat{\boldsymbol{e}}_1 + 0.7066 \hat{\boldsymbol{e}}_3 - 0.07606 \hat{\boldsymbol{e}}_4$, where $\hat{\boldsymbol{e}}_j$ is the $j^\text{th}$ unit vector, and discard the first 200 time units, approximately 3 Lyapunov times (LTs) \citep{racca_data-driven_2022}. The trajectories display a chaotic transient before relaminarisation occurs, in which the system reaches the equilibrium point given by $a_\text{eq} = \hat{\boldsymbol{e}}_1$.

\subsection{SDP-based analysis of Chaos Synchronisation}

For notational convenience, we largely consider the case where a given number, $p$, of the original system states are directly observed, i.e., $y = Ca = (a_i)_{i=1}^p$, with $C = \begin{pmatrix} I_p & 0_{p \times (n-p) }\end{pmatrix}$. Using the notation of \S \ref{sec:transformation} and the corresponding unobserved states are $\tilde{C}a = (a_i)_{i=p+1}^n$, the transformation $T=I$ is trivial. In this case $e=\hat{a}-a$ is the full-state error, and  $\tilde{e} = \tilde{C}e$ is the error in the unobserved modes only. 

Empirical numerical experiments suggest that for $p \geq 5$ pure synchronisation occurs, whereas it is possible to find initial conditions under which synchronisation fails for $p \leq 4$.  It is our aim now to determine if the SDP methods developed in \S \ref{sec:general_equiv_err_rom} can predict this behaviour, either directly from the governing equations or by exploiting statistical observations from the open-loop dynamics.

\subsubsection{An analytical study of pure synchronisation} \label{sec:diagonal_P}

We begin by discussing what can be proven analytically, by searching for diagonal matrices $P$ which solve \eqref{eq:pure_SDP}, since this will give important insights into the intrinsic challenges of practically verifying chaos synchronisation. We start by considering the case where $p=7$ modes are observed. On the sliding surface, the equivalent error dynamics are
\[
\dot{\tilde{e}} = A_{22}(a+\hat{a})\tilde{e}, \quad \text{with} \quad A_{22} = a_6 \begin{pmatrix} -\frac{c_{11}}{\text{Re}} & 0 \\ c_{12} & -\frac{c_{22}}{\text{Re}} \end{pmatrix},  
\]
where $c_{ij}$ are constants, depending only on the geometry of the flow domain. Since $a_6$ is bounded (we observe that $\|a_6\|_\infty \approx 0.5205$ for any solution) and the matrix $A_{22}$ is stable for any such $a_6$, it follows that the equivalent error dynamics are themselves stable, i.e. that $\lim_{t \rightarrow \infty} \tilde{e} = 0$ and thus pure synchronisation must occur.

While this analysis agrees with our empirical observations, it also highlights a crucial challenge in analysis this chaotic system.  At high Reynolds numbers (and even at the relatively moderate value $\text{Re}=400$ considered here) the matrix $A_{22}$ is very lightly damped, and is highly non-normal due to the non-zero off-diagonal entry $c_{12}$. As a consequence, small initial errors $\tilde{e}(0)$ can exhibit significant `transient growth' before eventually decaying to zero. 

This behaviour can be explained by our sufficient SDP condition \eqref{eq:pure_SDP} for chaos synchronisation. If stability is verified by finding a feasible solution $PA_{22}(a_6) + A_{22}(a_6) P^\top \prec 0$, classical Lyapunov theory can be used to describe the geometry of possible error trajectories in terms of properties of $P$. In particular, the `level sets' $\Omega_r = \{ x \in \mathbb{R}^n : x^\top P x \leq r \}$ are trapping regions for trajectories of $\tilde{e}(t)$ in the sense that once the error belongs to a level set of level $r$, it must remain within it for all subsequent times. `Transient growth' can only occur if the ellipsoidal level sets $\Omega_r$ are highly `stretched' and contain points both close to the origin and far from it; since, in this case, a small initial error $\tilde{e}(0)$ can take a large excursion in state-space before converging to zero.   

Mathematically, `highly stretched' corresponds to a large condition number of the matrix $P$. For this particular example, the stability condition $PA_{22}(a_6) + A_{22}(a_6)^\top P \prec 0$ can be satisfied for a diagonal $P=  \text{diag}(p,q)$ with $p>q>0$, and a simple calculation shows that the condition number $\kappa(P) = p/q$ must be of the order  $\kappa(P) \sim \|a_6\|_\infty^2 \text{Re}^2$---we find the optimal diagonal choice at $p=7, Re =400$ has $\kappa \approx 130.7$. Thus, transient growth of the error dynamics is likely to be observed, and this pathology gets worse with increasing Reynolds number. 

While this behaviour is not a barrier to analysis, it does highlight a key practical difficulty that must be overcome numerically: if any feasible solution to \eqref{eq:pure_SDP} has a high condition number, then the numerical implementation of the SDP is much more sensitive. Increasing the Reynolds number makes this problem worse, but observing fewer states has an even more detrimental effect. When $p=6$, for example, the equivalent error dynamics are governed by 
\[
A_{22}(a+\hat{a}) = a_6 \begin{pmatrix} -\frac{c_{11}}{Re} & 0 & -c_{13} \\ 0 & -\frac{c_{22}}{Re} & 0 \\ 0 & -c_{31} & -\frac{c_{33}}{Re} \end{pmatrix},
\]
for some constants $c_{ij}>0$. It is still possible to find a diagonal $P$ which \eqref{eq:pure_SDP}, but now its condition number must scale as $\kappa(P) \sim \|a_6\|_\infty^4  \text{Re}^4$ and we find the optimal diagonal choice for $p=6, Re=400$ has $\kappa \approx 2859$. 

An important question is therefore if there are ways to verify synchronisation, or design observers for assisted synchronisation, which avoids such ill-conditioned SDPs. For the cases described above, this is already hard. To analyse the case of observing only the first $5$ modes---where pure synchroniation is still observed to hold---is much more challenging since now $A_{22}$ depends on multiple system states $a_1,a_6,a_7,a_8$ and $a_9$. To make progress, we must move beyond simple analytical constructions and use semidefinite programming (SDP) to systematically design sliding mode observers for analysing chaos synchronisation.

\subsection{Synchronisation analysis using SDPs}

 We follow the approach of \S \ref{sec:general_equiv_err_rom} and seek to find feasible solutions $P$ to the SDPs \eqref{eq:pure_SDP} or \eqref{eq:pure_SDP_R}, but with `check-points' chosen to allow scalability to  higher-dimensional problems. 

As a preliminary step, we collect an ensemble of $N_b$ snapshots of the system state $(a(t_i))_{i=1}^{N_b} \subset \mathbb{R}^n$ and, using these, compute the  ensemble mean $\bar{a} \in \mathbb{R}^n$ and root-mean-square values $\gamma_\text{rms} \in \mathbb{R}^n$. A natural hierarchy of sets can be created by considering all check points within specified distances of the ensemble mean. Letting $R = (R_j)_{j=1}^n$ be some chosen component-wise distances, we define a set 
\begin{equation*}
\begin{split}
&S_R = \{ 2 \times a(t_i) : \; \left| \bar{a}_j - a_j(t_i) \right| \leq R_j, \; \forall j\}  \subset \mathbb{R}^n.
\end{split}
\end{equation*}
In general, the indices $j$ could cover all the modes, i.e., $j = 1, 2, \dots, n$. Considering only the modes that are active in the error dynamics, e.g., $a_6$ when $p = 7$ or $p = 6$, makes the problem less conservative. Recall that if \eqref{eq:pure_SDP} has a feasible solution with `check-points' belonging to $S_R$, then the Lyapunov function $V(\tilde{e})=\tilde{e}^\top P\tilde{e}$ for the error dynamics satisfies the key decay condition $\dot{V}<0$  whenever $a(t) + \hat{a}(t) \in \text{conv}(S_R)$. Thus, if a feasible solution can be found for sufficiently large $R$, chaos synchronisation is guaranteed. However, this approach also allows a more pragmatic study of convergence, in the sense that SDP feasibility for larger values of $R$ indicates an increased {\em likelihood} of convergence/synchronisation. It is therefore of interest to determine if there is a direct relationship between empirical observations of when synchronisation occurs, and SDP feasibility. 

To study this question, we consider three different choices: using the entry-wise root mean square (RMS) values $R_j= (\gamma_\text{rms})_j$; by using twice these values $R_j= 2(\gamma_\text{rms})_j$; or by using the maximum sampled values $R_j = \max_t |\bar{a}_j - a_j(t)|$. We denote the respective sets of check points $S_\text{rms}, S_{2 \text{rms}}$ and $S_{\max}$ and, if feasible solutions to \eqref{eq:pure_SDP} or \eqref{eq:pure_SDP_R} can be found we denote these $P_\text{rms}^\ast, P_{2\text{rms}}^\ast$ or $P_\text{max}^\ast$. These SDPs are posed with YALMIP \citep{lofberg_yalmip_2004} and solved with MOSEK 11.2 \citep{mosek_aps_mosek_2026} on MATLAB. For numerical stability, we impose the sampled semidefinite constraints with  a small tolerance $PA_{22}(s_i) + A_{22}(s_i)^\top P \preceq -\epsilon I$, using $\epsilon = 10^{-3}$ throughout. Check points are drawn from $N_b = 2,000$ samples of the system trajectory.  Finally, due the possibility of large condition numbers noted in \S \ref{sec:diagonal_P}, in every case we add the extra constraint $I_{n-p} \preceq P \preceq \kappa I_{n-p}$ and search for a matrix $P$ which minimizes the condition number bound $\kappa$ \citep{boyd_linear_1994}. 

\begin{table*}[h]
\centering
\caption{Condition number $\kappa(P)$ of feasible solution to the pure synchronisation design problem \eqref{eq:pure_SDP} and the assised synchronisation design problem \eqref{eq:pure_SDP_R} for different choices of check points set $S_R$. Entries are marked `$\dagger$' where no feasible solution could be found. Also indicated in the second column is whether synchronisation has been attained empirically in simulations.}
\vskip 1em 
\label{tab:cond_P}
\begin{tabular}{c ccccccccc}
\toprule
& & \multicolumn{8}{c}{Condition number $\kappa(P)$ \rule{2.5em}{0pt}} \\
\cmidrule(lr){3-10}
& & \multicolumn{4}{c}{Pure Synchronisation \eqref{eq:pure_SDP}} & \multicolumn{4}{c}{Assisted Synchronisation \eqref{eq:pure_SDP_R}} \\
\cmidrule(lr){3-6} \cmidrule(lr){7-10}
\multirow{2}{6em}{\# Modes Observed ($p$)} & \multirow{2}{4em}{Empirical synch.} & $P_{\text{rms}}^\ast$ & $P_{\text{2rms}}^\ast$ & $P_{\text{max}}^\ast$ & $P_{\text{diag}}$ & $P_{\text{rms}}^\ast$ & $P_{\text{2rms}}^\ast$ & $P_{\text{max}}^\ast$ & $P_{\text{diag}}$ \\
&  &   &  &  &  &  &    &  & \\
\midrule
7 & \checkmark & 5.469 & 21.50 & 92.75 & 130.7 & 1 & 1 & 1 & 1 \\
6 & \checkmark & 7.787 & 82.23 & 2504 & 2859 & 1 & 1 & 3.182 & 4.165\\
5 & \checkmark & 40.73 & 244.5 & $\dagger$ & $\dagger$ & 30.89 & 58.51 & 68.04 & 1380 \\
4 & $\times$ & $\dagger$ & $\dagger$ & $\dagger$ & $\dagger$ &  $\dagger$ & $\dagger$  & $\dagger$ & $\dagger$ \\
\bottomrule
\end{tabular}
\end{table*}

Table \ref{tab:cond_P} shows the condition numbers of the optimal decision variables $P$, for the observer design problem using different numbers of observed modes, and different choices of $S_R$. The flexibility to search for non-diagonal solutions to \eqref{eq:pure_SDP} can be seen to improve the condition number of $P_{\max}^\ast$, over and above diagonal solutions, by 30\% and 12\% when $p=6$ and $7$, respectively. More importantly, assisted synchronisation {\em significantly} improves the condition numbers of the optimal SDP solutions. Further, it is required to find feasible solutions when using the full check point set $S_{\text{max}}$ in the case $p=5$. Interestingly, the switch from feasibility to infeasibility between $p=5$ and $p=4$ is not preceded by a significant increase in optimal condition number when using assisted synchronisation. This strongly suggests that feasibility of \eqref{eq:pure_SDP_R} analytically captures the empirical observation that five observed modes is the `natural' boundary, at least for  this example, of where chaos synchronisation occurs.

\begin{figure}[tb]
    \includegraphics[width=\linewidth]{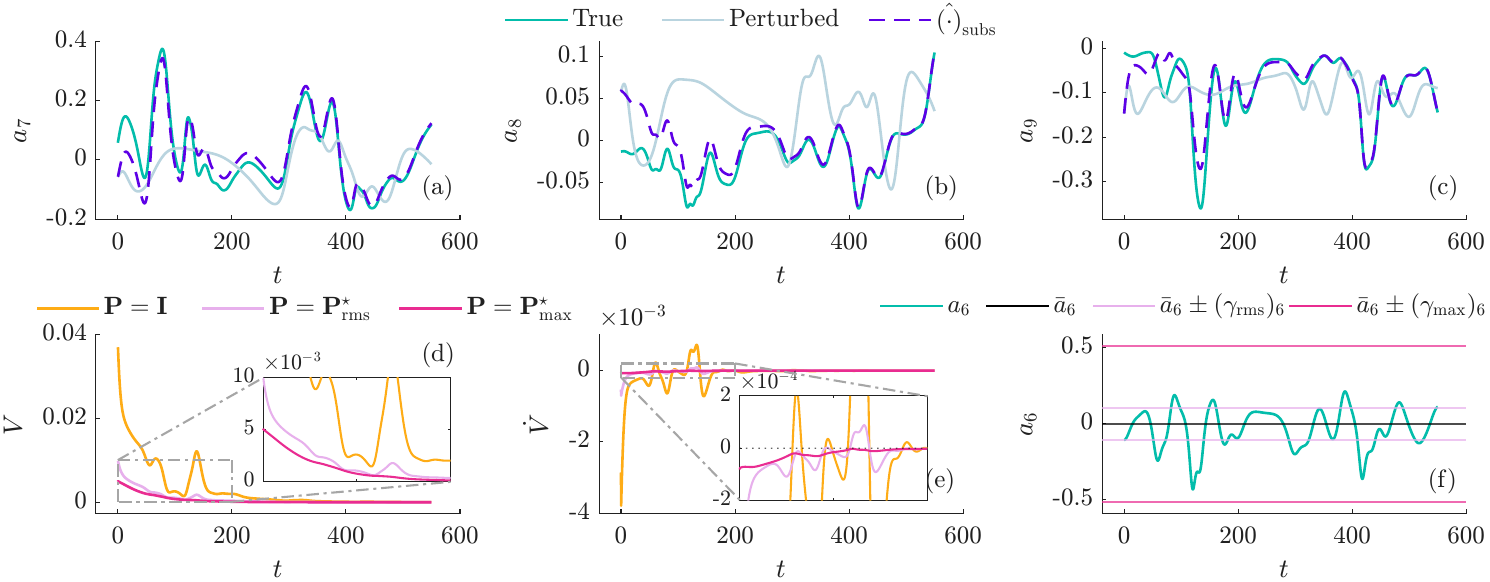}
    \caption{Pure synchronisation of the unobserved modes when observing the first six modes of the nine-dimensional model of turbulent shear flow (a -- c).  Lyapunov functions (d) and their time derivatives (e) as a function of time for different choices of $P$. The validity of the chosen $P$ depends on the perturbations of $a_6$ (f).}
    \label{fig:mfe_c6_lyap}
\end{figure}

Figure \ref{fig:mfe_c6_lyap} exemplifies the relation between chaos synchronisation and the feasibility of the SDP \eqref{eq:pure_SDP}, in the case $p=6$.  Trajectories of the unobserved states $(a_i)_{i=7}^9$ are shown in Figure \ref{fig:mfe_c6_lyap}~(a-c) for both the true trajectory, and the one arising from direct substitution initialised using a small perturbation to the true state's initial condition (we use a random perturbation $\hat{a}(0) = a(0) + \epsilon \hat{\boldsymbol{e}}_4$ with $\epsilon \sim \mathcal{U}(0, 0.05)$). This perturbed initial condition emphasises the chaotic nature of the dynamics in the sense that a small perturbation gives rise to a significantly different trajectory. Starting from the shown time window, we feed observations to the slave system, and after an initial transient, it is clear that synchronisation occurs. The chosen window exemplifies the worst case scenario for provably verifying pure synchronisation \eqref{eq:pure_SDP}, as $|a_6(t)|$ comes close to the maximum value we observe in all simulation data---see Figure \ref{fig:mfe_c6_lyap} (f) at $t \approx 130$. 

Lyapunov functions $V(\tilde{e}(t)) = \tilde{e}(t)^\top P\tilde{e}(t)$ for different choices of $P$, and their time derivatives, are shown in Figure \ref{fig:mfe_c6_lyap}~(d-e). These allow us to interrogate whether the SDP approach can effectively describe or explain the observed synchronisation behaviour. Recall that it is only if we observe a monotonically decreasing trajectory of $V(\tilde{e}(t))$ that the constructed Lyapunov function can be a potential candidate for explaining synchronisation. 

A preliminary observation is that analysing the `energy' of the error $V_I(\tilde{e}(t)) = \| \tilde{e}(t)\|^2 = \tilde{e}(t)^\top I\tilde{e}(t)$ cannot explain synchronisation, since its trajectory is clearly non-monotonic (yellow lines in Figure \ref{fig:mfe_c6_lyap} (d -- e)). If the SDP \eqref{eq:pure_SDP} is solved using the check points $R_\text{rms}$, then the behaviour of the trajectory of the corresponding Lyapunov candidate $V_\text{rms}(\tilde{e}) = \tilde{e}^\top P^\ast_{\text{rms}} \tilde{e}$ is markedly improved.  Its trajectories, shown in pink, are close to being monotonic apart from within some short time intervals, e.g., around $t \approx 130$, at which the state $a_6$ takes its largest excursion from the its mean value. Finally, the feasible solution $P^\ast_{\text{max}}$ using check points $S_{\text{max}}$ can be seen in Figure \ref{fig:mfe_c6_lyap}~(f) to consider all observed values of the crucial state $a_6$ and, when this is achieved, the corresponding Lyapunov function decays monotonically. This confirms the analytical proof of convergence using a diagonal $P$ from \S \ref{sec:diagonal_P}, in addition to demonstrating the influence of the size of $S_R$ on the monotonicity properties of the corresponding Lyapunov functions. 

We now consider the more challenging case of observing $p=5$ modes---which is the empirically-observed limit of when pure synchronisation occurs. As reported in Table \ref{tab:cond_P}, no feasible solution to the pure synchronisation SDP \eqref{eq:pure_SDP} can be found if the full check points set $S_{\max}$ is used. However, feasible SDPs can be constructed, either by reducing the number of check points or by using assisted synchronisation. The smaller check points sets $S_\text{rms}$ and $S_{2 \text{rms}}$ contain $28.2\%$ and $81.6\%$ of the original $N_b = 2,000$ samples, respectively, and the SDP \eqref{eq:pure_SDP} is feasible in both cases yielding solutions
\begin{equation*}
    P^\ast_{\mathrm{rms}} = \begin{bmatrix}
    1.1018 &  0.0902  & 0.7079 & -0.0837 \\
    0.0902 &  1.4164  & 4.0154 &  0.0039 \\
    0.7079 &  4.0154  & 40.3093 &  0.0952 \\
   -0.0837 &  0.0039  & 0.0952  &  1.1317
    \end{bmatrix},
\end{equation*}
and
\begin{equation*}
    P^\ast_{\mathrm{2rms}} = \begin{bmatrix}
    1.0458 &  -0.0122  &  -0.7163 &  0.0866 \\
   -0.0122 &   1.2050  &   6.9613 &  0.0558 \\
   -0.7163 &   6.9613  & 244.2494 & -0.0419 \\
    0.0866 &   0.0558  &  -0.0419 &  1.5632 
    \end{bmatrix},
\end{equation*}
with $\kappa^\ast = 40.73$ and $\kappa^\ast = 244.5$, respectively. Table \ref{tab:cond_P} shows that these condition numbers are significantly higher than in the cases $p=6,7$, as predicted by the analysis in \S \ref{sec:diagonal_P}.  We find that the boundary of feasibility of the SDP is using $S_R$ containing points within 2.27 standard deviations from the mean ($87.5\%$ of the original samples),  at which point the optimal condition number becomes $\kappa^\ast = 912.2$. 

\begin{figure}[tb]
    \includegraphics[width=\linewidth]{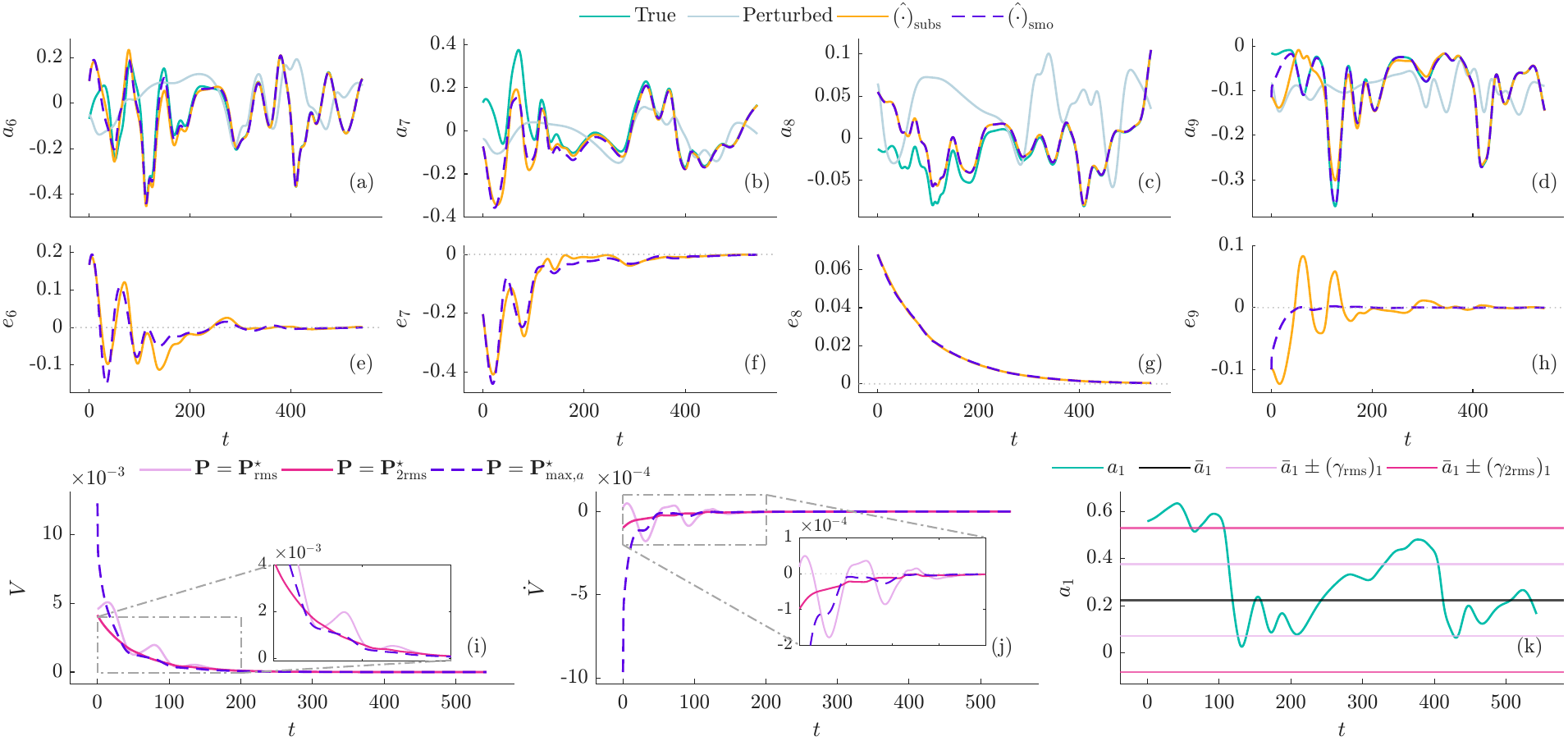}
    \caption{Pure and assisted synchronisation of the unobserved modes when observing the first five modes of the nine-dimensional model of turbulent shear flow (a -- d). Lyapunov functions (i) and their time derivatives (j) as a function of time for different choices of $P$ with the subscript `a' denoting the assisted case. Shown is the time window after sliding is reached for both cases. The feasibility of the chosen $P$ depends on the perturbations of multiple state variables; $a_1$ (k), $a_6$, $a_7$, $a_8$, $a_9$.}
    \label{fig:mfe_c5_lyap}
\end{figure}

Figure \ref{fig:mfe_c5_lyap} shows indicative trajectories of the flow and observers for the case $p=5$. Note that synchronisation still occurs. Monotonicity of the constructed Lyapunov functions is now much harder to achieve, with Figure \ref{fig:mfe_c5_lyap}~(i -- j) indicating that, in the case of pure synchronisation analysis, using the larger constraint set $S_{2 \text{rms}}$ is necessary for monotonicity. In the analysis of the assisted case, first, we set the sliding feedback gains as $K_s = 0.05I_5$, which quickly leads to sliding in the observed modes in about 8 time units. The assisted feedback gains $K_a$ enable the feasibility check with the maximum constraint set $S_{\text{max}}$, yielding the optimal design variables
\begin{equation*}
    P_{\text{max}}^\ast = \begin{bmatrix}
    1.6364  &  0.0979 &   0.4902  & -0.1217 \\
    0.0979  & 1.6868  &   6.7322  &  0.0665 \\
    0.4902  &  6.7322 &   67.3471 &  0.2826 \\
   -0.1217  &  0.0665 &   0.2826  & 59.7466 
    \end{bmatrix},
\end{equation*}
and
\begin{equation*}
    K_a^\ast = 10^{-4} \times \begin{bmatrix}
    38   &   1  &  0 &  0 & -1 \\
    22   &   0  & -3 &  6 & 21 \\
    -4   &   0  &  2 &  1 & -2 \\
    486  &   8  &  5 &  0 & 0
    \end{bmatrix},
\end{equation*}
with a condition number of $\kappa^\ast = 68.04$, which is significantly lower compared to the condition numbers obtained with pure synchronisation. Inspecting the gain matrix, we expect that the errors in $a_6$, $a_7$ and mostly $a_9$ being controllable by feedback, whereas the linear damping in $a_8$ sets an upper bound to the maximum achievable exponential decay rate, which can be seen in Figure \ref{fig:mfe_c5_lyap}~(e -- h). Once sliding occurs, the time derivative of the designed Lyapunov function (dashed line in Figure \ref{fig:mfe_c5_lyap}~(j)) remains below zero in line with our stability objective, illustrating the efficacy of the approach in both explaining and making improvements upon pure synchronisation.
%that using the larger check point set $P_{2 \text{rms}}$ is necessary to ensure monotonicity of associated trajectories $V(\tilde{e}(t))$. 

%where the time derivative of the Lyapunov function, $\dot{V}$ associated with $\matr{P}_{2rms}$ remains below zero for the majority of the considered trajectory, except for a few time steps around $t = 1380$, which is possibly a combination of the effects of the individual mode perturbations. 

When only the first four modes, $Ca = (a_1, a_2, a_3, a_4)^\top$, or another combination of four modes such as, $Ca = (a_1, a_2, a_3, a_5)^\top$, are observed, we do not obtain synchronisation of the unobserved modes. Feedback has the potential to stabilise the error dynamics of the unobserved modes and make chaos synchronisation possible. However, for the mentioned cases, we have been unable to find feasible stabilising solutions using the SDP approach. 

\section{Transformation to the companion form via non-linear observability}\label{sec:nonlinear_observability}
\begin{figure}[tb]
    \includegraphics[width=0.95\linewidth]{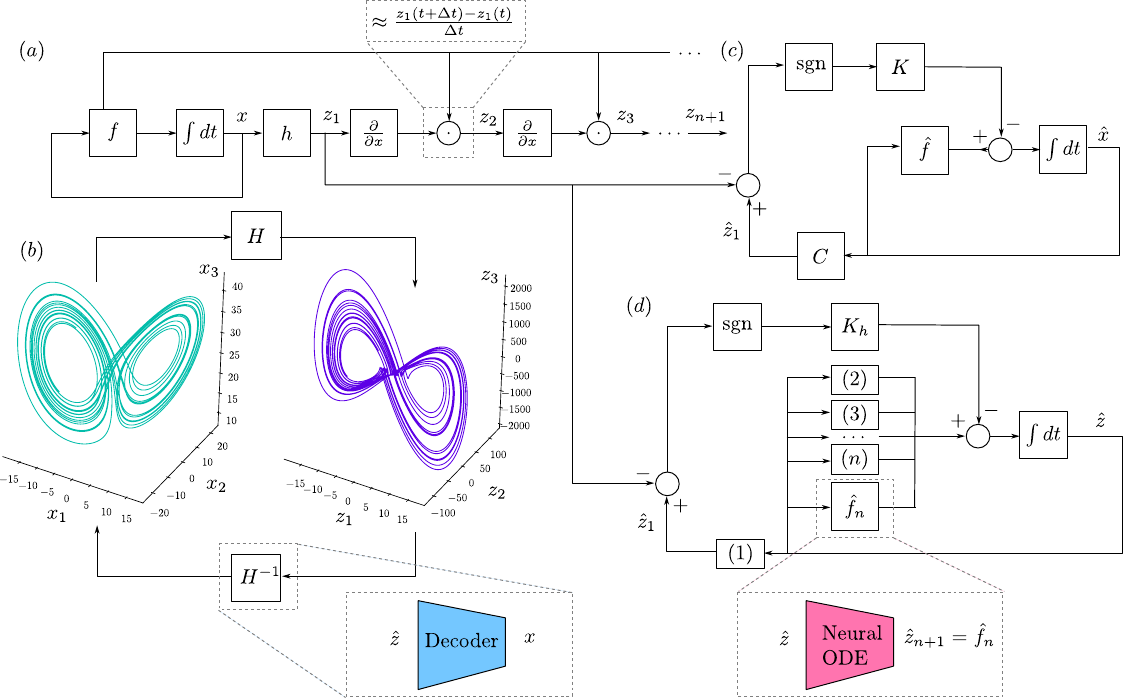}
    \caption{(a) True dynamics $\dot{x} = f(x)$ and the construction of Lie derivatives of the observation operator $h$. (b) The Lie differentiation represents a transformation $H$, and if the system is non-linearly observable, there exists an inverse transformation $H^{-1}$. The decoder's goal is to learn this inverse transformation from data. (c) Sliding mode observer in the physical space with the feedback gains $K$. (d) Sliding mode observer in the transformed space with the feedback gains $K_h$. The neural ODE approximates the companion form dynamics in this space.}
    \label{fig:transformation_schematic}
\end{figure}
For known reduced-order models, the SDP approach offers a way to determine observer gains for any given linear observation matrix. However, this process relies on (i) the functional form of the searched Lyapunov function, here we used an energy-like quadratic prior, (ii) the functional form of the observer gains, here we used a time-invariant first-order feedback, and (iii) checking the LMI conditions at the parameter vertices, or alternatively, sampling from the chaotic attractor. First, it may be impossible to find feasible Lyapunov functions or gains of the form we have set a priori. Second, the number of parameter dependencies and the state dimension might make the solution of the SDP computationally intractable. In this section, we investigate an alternative approach based on the companion form of the sliding mode observer \eqref{eq:companion_form}, in which the observer design reduces to setting the poles of a linear system. It may be possible to analytically derive the companion form equations through the definition of non-linear observability introduced in \eqref{eq:nl_observability}. We first demonstrate this on the Lorenz system, highlighting two challenges that arise: the complex non-linearities that emerge even for this low dimensional system, and the existence of an inverse map to the original state space. These challenges motivate the learning-based approach proposed in \S\ref{sec:neural_network}, which in turn enables the design of convergent observers for configurations that have previously proved to be infeasible using the SDP approach, e.g., $p = 4$ in Table \ref{tab:cond_P}. The robustness of the sliding mode observers to uncertainties \citep{slotine_sliding_1986} make them especially suitable for counteracting approximation errors of the learning-based approach.

\subsection{Non-linear observability}
Consider the non-linear system transformation \eqref{eq:transformation} with the variables $z = \left(z_i\right)_{i=1}^{n} = H(x)$ with $z_1=h(x) \in \mathbb{R}$ and 
\begin{equation}
    z_{i+1} =  \frac{\partial z_i}{\partial x}f(x), \quad i = 1, 2, \dots, n-1,
\end{equation}
comprised of successive Lie derivatives of the observed states $h(x)$ along the trajectories of system \eqref{eq:nonlin_sys}. Because $x$ is a solution of the non-linear system \eqref{eq:nonlin_sys}, $z_i$ is the $(i - 1)^{\text{th}}$ order time derivative of the observation $y$ and the unknown dynamics in the companion form \eqref{eq:companion_form} is consequently given by \citep{zeitz_observability_1984}
\begin{equation}
    f_n(z) = \frac{\partial z_n}{\partial x} f(x)\vert_{x = H^{-1}(z)}.
\end{equation} 
If the map $H$ is an injection, then designing an observer such that the error dynamics of $z$ \eqref{eq:companion_form_error} are stable guarantees the stability in the error dynamics of the original space $x$, as one can deploy the inverse map $H^{-1}$ to reconstruct $x$ from $z$. To see this, linearising $H$ around $x = \hat{x}$, we get
\begin{equation}\label{eq:drakunov}
    x = \hat{x} - \left(\left.\frac{\partial H}{\partial x}\right\vert_{x = \hat{x}}\right)^{-1} (\hat{z} - z),
\end{equation}
which retrieves local observability condition \eqref{eq:nl_observability} based on the rank of the Jacobian, i.e., the system is locally observable if the Jacobian is full rank (and is therefore invertible). \citet{drakunov_sliding_1995} propose a sliding mode observer which relies on the local inversion \eqref{eq:drakunov}. Although this observer is theoretically guaranteed to converge with the appropriate selection of the gains, we find that it is practically difficult to implement, as it requires for the inverse of the Jacobian to be defined at all times. To circumvent this practical difficulty, we take an alternative approach such that the sliding mode observer operates fully in the transformed space and the original state is only recovered in a post-processing step using either an analytical expression or a data-driven approximation of $H^{-1}$.  
\subsection{Companion form of the Lorenz system}\label{sec:lorenz_companion}
The concepts of state transformation and non-linear observability can be easily illustrated on the Lorenz system. Considering the case of $x_1$ observations, the transformations are given by the maps
\begin{equation}
    H = \begin{pmatrix}
        x_1 \\
        \sigma(x_2-x_1) \\
        -\sigma^2 (x_2-x_1) + \sigma(x_1(\rho-x_3) - x_2)
    \end{pmatrix}, \label{eq:lorenz_transform}
\end{equation}
and,
\begin{equation}
    H^{-1} = \begin{pmatrix}
        z_1 \\
        z_1 + z_2/\sigma \\
        \left(-z_3 - z_2(\sigma + 1) + \sigma z_1(\rho-1)\right)/\sigma z_1
    \end{pmatrix}, \label{eq:lorenz_inv_transform}
\end{equation}
where the non-linear observability matrix is given by
\begin{equation}\label{eq:lorenz_observability}
    \frac{\partial H}{\partial x} = \begin{pmatrix}
        1 & 0 & 0 \\
        -\sigma & \sigma & 0 \\
        \sigma^2 + \sigma (\rho - x_3) & -\sigma^2 -\sigma & -\sigma x_1
    \end{pmatrix}.
\end{equation}
Notice that the observability matrix loses rank in the last column, i.e., in $x_3$,  when $x_1 = 0$. However, $\dot{x}_3 = -\beta x_3$ when $x_1 = 0$, which is stable. Therefore, the system remains detectable, as the unobservable state is stable. 

In linear systems, the order of differentiation in the observability matrix is limited due to the Cayley-Hamilton theorem, as differentiating the system dynamics further does not increase the rank. This limit does not necessarily apply to non-linear systems. In fact, for the Lorenz system including the third Lie derivative, the system becomes locally unobservable for $x_3$ if $x_1 = x_2 = 0$. The remarks regarding observability appear to have a strong link to the Takens embeddings of chaotic systems, and the analysis of \citet{cross_differential_2010}, who show that the differential map with three dimensions is not an embedding of the Lorenz system, whereas four is.

Utilising the inverse transformation \eqref{eq:lorenz_inv_transform}, the sought dynamics for the final state variable in the transformed space is
\begin{equation}\label{eq:lorenz_z3_dot}
\begin{split}
    \dot{z}_3 =&-(1 + \beta + \sigma)z_3
               + \beta (1 + \sigma)z_2 
               + z_1^2 z_2 \\
               &\quad + \sigma z_1^3
               + \beta \sigma (1 - \rho) z_1
            +\frac{z_2}{z_1} (z_3 + z_2 (1 + \sigma)),
\end{split}
\end{equation}
which we can then use in the design of a sliding mode observer. Notice the complexity of this relationship even for the modest three dimensional system with quadratic non-linearity. The dynamics \eqref{eq:lorenz_z3_dot}, and similarly the inverse transformation \eqref{eq:lorenz_inv_transform} would become undefined when $x_1 = z_1 = 0$ due to the $1/z_1$ term, which coincides with the condition that the local observability matrix \eqref{eq:lorenz_observability} loses rank. However, in this case substituting  the definitions of $z_2$ and $z_3$ from \eqref{eq:lorenz_transform} with $x_1 = 0$, this last term vanishes and the companion form remains well behaved \citep{cross_differential_2010}. Nonetheless, the loss of observability indicates that numerical issues may arise when approximating these functions from data. Analysing observations of $x_2$ instead, in which case synchronisation also occurs, the inverse transformation map with $z_1 = x_2$ %is
%transformation map is
%\begin{equation}
%    H = \begin{pmatrix} 
%        x_2, \\
%        x_1(\rho-x_3)-x_2,\\
%        x_2-x_1(\rho-x_3)+x_1(\beta x_3 - x_1x_2)-\sigma(\rho-x_3)(x_1-x_2),\\
%    \end{pmatrix},
%\end{equation}
%\begin{equation}
%    \begin{split}
%  &-z_1 x_1^3+\beta\rho x_1^2-(z_3+z_2+(\beta+\sigma)(z_2+z_1))x\\ &\quad +\sigma(z_2+z_1)z_1 = 0,\\
%    & x_3 = \rho-\frac{z_2+z_1}{x_1},
%    \end{split}
%\end{equation}
has three roots for the solution of $x_1$, and consequently $x_3$, and the map is therefore not an injection. However, numerically computing these roots, we find that including the third Lie derivative $z_4$ provides the necessary information to determine $x_1$. When observing $x_3$ on the other hand, the synchronisation does not occur and our analysis shows that in this case the inverse map has four roots. Some of these roots can be discarded as they are not on the chaotic attractor; or can be distinguished via the third Lie derivative; or are imaginary. However, the symmetric pairs, $(x_1,x_2)$ and $(-x_1, -x_2)$ remain indistinguishable from the derivatives of $x_3$ alone. 

The analysis of observability helps us understand the conditions under which we expect synchronisation to occur and can reliably work with the companion form. It also highlights how complex these relationships can be to determine analytically, which motivates our learning-based approach introduced next.
%The nonlinear observability matrix is
%\begin{equation}
%    \frac{\partial \vect{H}}{\partial \vect{x}} = \begin{bmatrix}
%        0 & 1 & 0 \\
%        \rho - z&  -1 &  -x\\
%        \beta z - 2xy - (\sigma+1)(\rho - z)& - x^2 + \sigma(\rho - z) + 1& (\beta + 1) x + \sigma(x - y)\\
%    \end{bmatrix}.
%end{equation}
%Finally, for $x_3$ observations, the inverse transformation map with $x_3 = z_1$ is
% \begin{align}
%    h_1 &= z,\\
%    h_2 &= xy-\beta z,\\
%    h_3 &= \beta(\beta z-xy)-x(y-x(\rho-z))-\sigma y(x-y)
%\end{align}
%\begin{equation}
%    \begin{split}
%        &(\rho-z_1)x_1^4+(\beta^2 z_1 - (\beta+\sigma+1)z_2+\beta z_1 -%z_3)x_1^2\\ & \quad +\sigma(z_2+\beta z_1)^2 = 0,\\
%        &x_2 = \frac{z_2+\beta z_1}{x_1},
%    \end{split}
%\end{equation}
%in which case the solution of $x_1$, and consequently $x_2$, has four roots, and we find that including the information from further Lie derivatives does not help determining $x_1$. This analysis confirms our empirical observations of the lack of synchronisation with $x_3$ observations.
%The nonlinear observability matrix is
%\begin{equation}
%    \frac{\partial \vect{H}}{\partial \vect{x}} = \begin{bmatrix}
%        0 & 0 & 1 \\
%        y &  x &  -\beta\\
%        2x(\rho - z)-(1+\sigma+\beta)y& 2\sigma y - (1+\sigma+\beta)x& \beta^2 - x^2\\
%    \end{bmatrix}.
%\end{equation} \\
% include when it becomes locally unobservable
\section{A neural network approach to the sliding mode observer}\label{sec:neural_network}
For spatiotemporally chaotic systems such as turbulent flows that are high-dimensional, the transformation to the companion form \eqref{eq:companion_form} may be unknown or analytically intractable. We propose a fully data-driven approach to approximate (i) the state dynamics in this transformed space for the design of a sliding mode observer \eqref{eq:sliding_observer_companion}; and (ii) the mapping from this space to the original physical space, which we are interested in estimating. This method is visualised in Figure \ref{fig:transformation_schematic} (b) and (d), while the process of  Lie differentiation is shown in (a) and the general sliding mode observer with linear observation matrix is shown in (c).

The proposed architecture has two components. The first component is a neural ordinary differential equation (ODE), which learns a mapping from the observations $y = (y_i)_{i = 1}^{p}$ and their time derivatives up to order $m - 1$, $(\dot{y}_i, \ddot{y}_i, \dots, y_i^{(m-1)})_{i = 1}^{p}$, to the time derivative of the final state variable $(y_i^m)_{i=1}^{p}$. Extending the companion form \eqref{eq:companion_form} to multiple observations, this means predicting
\begin{equation}\label{eq:neural_ode}
    \hat{Z}_{m+1} = \hat{f}_m(Z;\theta_f),
\end{equation}
where $Z = (Z_1^\top \; Z_2^\top \; \dots \; Z_{m}^\top)^\top$ is the concatenated input with $Z_{j} \in \mathbb{R}^{p}$ denoting the $(j-1)^{\text{th}}$ Lie derivative of $h(x)$, and $f_m: \mathbb{R}^{p \times m} \times \mathbb{R}^{n_f} \rightarrow \mathbb{R}^{p}$ is a neural network that approximates the dynamics $f_m$ with $\theta_f \in \mathbb{R}^{n_f}$ representing its trainable weights. If we consider a single observation $p = 1$, then we expect $m = n$ the dimension of the original state space. For multiple observations $p > 1$, the linear part of the companion form dynamics \eqref{eq:companion_form} are decoupled and each block corresponding to an observation can be designed independently. As for the non-linear part, we exploit the flexibility of neural networks and allow for couplings between state variables stemming from different observations. Once the training is performed, the weights are fixed and the system of ODEs \eqref{eq:companion_form} can be time-marched autonomously with a given initial condition. In this case, the input to $\hat{f}_m$ becomes $\hat{Z}$ to denote the state estimate. The space of the observations and their Lie derivatives can be thought of as a latent space, where the dynamics are constrained to be in the staircase form of \eqref{eq:companion_form}. We implement the sliding mode observer in this space to correct the state estimate using the feedback from the observations. 

The second component of the proposed architecture is a decoder, which learns the inverse transformation from the state in the observer space, $Z$, to the state in the original physical space, $x$
\begin{equation}\label{eq:decoder}
    \hat{x} = \delta(Z; \theta_\delta),
\end{equation}
where $\delta: \mathbb{R}^{p \times m} \times \mathbb{R}^{n_\delta} \rightarrow \mathbb{R}^{n}$ is a neural network that approximates $H^{-1}$ with $\theta_\delta \in \mathbb{R}^{n_\delta}$ denoting its trainable weights. While the decoder is not necessary for the sliding mode observer \eqref{eq:companion_form}, its existence proves that the map $H$ is an injection, which allows us to perform the observer design in the latent space, and map the state estimate to the physical state. It is also noteworthy that the Jacobian of the decoder $\partial \delta/\partial Z$, which can be easily obtained through automatic differentiation, approximates the inverse of the local observability matrix $(\partial H/\partial x)^{-1}$ \eqref{eq:drakunov}. Furthermore, the decoder Jacobian also allows for transformation of the observer gains from one space to the other. If we denote the gains in the latent space as $K_h$, then the corresponding gains in the original space are $K = (\partial H/\partial x)^{-1}K_h$.

Given a dataset $\{\{Z^{r}_1(t_k), x^{r}(t_k)\}_{k=0}^{N-1}\}_{r=1}^{R}$, where $t_k = k\Delta t$ are the discretised time steps and $R$ is the number of trajectories with different initial conditions, we first compute $Z^{r}_j(t_k), \; j = 2, 3, \dots, m + 1$ using a numerical differentiation scheme. The order of differentiation $m$ relates to the required embedding dimension for the chaotic attractor and the available number of observations. However, practically, the maximum number of differentiations we can perform reliably is limited as each order of differentiation causes loss of information, with the amplitude of low frequency content ($\omega < 1$) tending to zero, and the amplitude of high frequency content ($\omega > 1$) tending to $\infty$. Therefore, we treat $m$ as a hyperparameter to be tuned. Following the generation of the input and output sets, we train the weights of the neural ODE and the decoder individually to minimise the respective losses using the mean square error (MSE)
\begin{equation}
    \theta_f^\ast
    = 
    \underset{\theta_f}{\arg\min} \;
    \mathcal{L}_f, 
    \quad 
    \theta_\delta^\ast
    = 
    \underset{\theta_f}{\arg\min} \; 
    \mathcal{L}_\delta,
\end{equation}
where
\begin{equation}
    \mathcal{L}_f 
    = 
    \frac{1}{N_\mathrm{tot}}
    \sum_{r=1}^{R}\sum_{k=m k_l}^{N-m k_r}
    \left\|Z^{r}_{m+1}(t_k)
    -
    \hat{f}_m(Z^{r}(t_k);\theta_f)
    \right\|_2^2,
    \label{eq:node_loss}
\end{equation}
\begin{equation}
    \mathcal{L}_\delta 
    = 
    \frac{1}{N_\mathrm{tot}}\sum_{r=1}^{R}
    \sum_{k=m k_l}^{N-m k_r}
    \left\|x^{r}(t_k)
    -
    \delta(Z^{r}(t_k);\theta_\delta)
    \right\|_2^2,
    \label{eq:decoder_loss}
\end{equation}
and $k_l$ and $k_r$ denote the number of time steps discarded from the data depending on the order of accuracy of the numerical differentiation scheme (e.g., for a first order scheme, $k_l = 0$ and $k_r = 1$), and $N_\mathrm{tot} = R(N-m(k_l+k_r))$ is the total number of points in the dataset after pre-processing.

Instead of the pointwise loss on the final time derivative \eqref{eq:node_loss}, the training of the neural ODE $\hat{f}_m$ can also be performed using a rollout loss, which evaluates the prediction error of time integrated trajectories
\begin{equation}
\label{eq:neural_ode_rollout_loss}
\begin{aligned}
    \mathcal{L}_{f}^{\mathrm{roll}}
    ={}&
    \sum_{r=1}^{R}
    \sum_{k=mk_l+1}^{N-mk_r-N_{\mathrm{w}}+1}
    \frac{1}{N_{\mathrm{w}}}
    \sum_{j=0}^{N_{\mathrm{w}}-1}
    \left\|
        Z^r(t_{k+j})-\hat{Z}^r(t_{k+j})
    \right\|_2^2 \\
    &\text{subject to} \\
    &\hat{Z}^r(t_{k+j})
    = \int_{t_k}^{t_{k+j}}
      \hat{f}\!\left(\hat{Z}^r(t);\theta_f\right)\,\mathrm{d}t,
    \\
    &\hat{Z}^r(t_k)=Z^r(t_k).
\end{aligned}
\end{equation}
where $N_\mathrm{w}$ is the number of time steps in one window of evaluation. In this training, the gradient of the weights with respect to the loss are calculated using the backward in time integration of the adjoint equations around the forward trajectory \citep{chen_neural_2018}. While this type of training promotes numerical stability of the integrated neural ODE, it is more computationally expensive due to back-propagation through time. Furthermore, for chaotic systems, the adjoint equations can only be integrated for short time windows before the gradients diverge exponentially. Therefore, we perform a first training of the weights using $\mathcal{L}_f$, and then the weights can be fine-tuned using $\mathcal{L}_{f}^\mathrm{roll}$ if needed. This step is necessary if we intend to use the learned model for forecasting or model predictive control purposes, which would require accurate integration of the ODE. For the current observer, the stability of the learned ODE is not critical, as the feedback from very frequently fed observations prevent the trajectories from diverging. 
\subsection{Results for the Lorenz system}
We apply the proposed neural network approach to the Lorenz system, for which we have analysed the transformations to the companion form in \S\ref{sec:lorenz_companion}. We generate a dataset of four trajectories of length $10^4$ time steps (with $\Delta t = 10^-3$ corresponding to approximately 11 Lyapunov times) by perturbing the initial condition $x_0 = (1, 0, 0)^\top$ with uniform sampling from $\mathcal{U}(-0.1,0.1)$ and discarding the first $2 \times 10^4$ time steps of integration. We set three of these trajectories for training and one for validation. The input to both the neural ODE and decoder networks is $Z = (z_1, z_2, z_3)^\top = (x_1, \dot{x}_1, \ddot{x}_1)^\top$. To keep the procedure fully data-driven, we approximate the time derivatives using a fourth-order numerical differentiation scheme. The ground truth labels for the neural ODE are $z_4 = x_1^{(3)}$, and for the decoder, they are the full state vectors $x = (x_1, x_2, x_3)^\top$. Before training, the inputs and the outputs are standardised to have zero mean and unit standard deviation. We construct both neural networks with two layers of 64 neurons each and with $\tanh$ as the activation function. The training is then performed in batches using gradient descent with the Adam optimizer at a learning rate of $ 10^{-4}$. A critical point in the training is the chosen batch size. Due to the peaks in the time derivatives, we require a batch size of one, i.e., the gradient step is applied per data point, as averaging over a batch smooths outs the prediction over the peaks. After 10 epochs of training, both networks achieve good accuracy and generalisation over all, except that the decoder has error in $x_3$ predictions when the trajectory crosses from the positive $x_1x_2$-plane to the negative $x_1x_2$-plane. (This switching from one unstable equilibrium to the other is typical for the Lorenz system.) Adding the third time derivative in the inputs does not alleviate this error. This finding is in line with our analysis of the loss of observability of $x_3$ for $x_1 = x_2 = 0$, and indicates that the observability condition puts a limitation on how well such decoders can be trained. 
\begin{figure}[tb]
    \includegraphics[width=\linewidth]{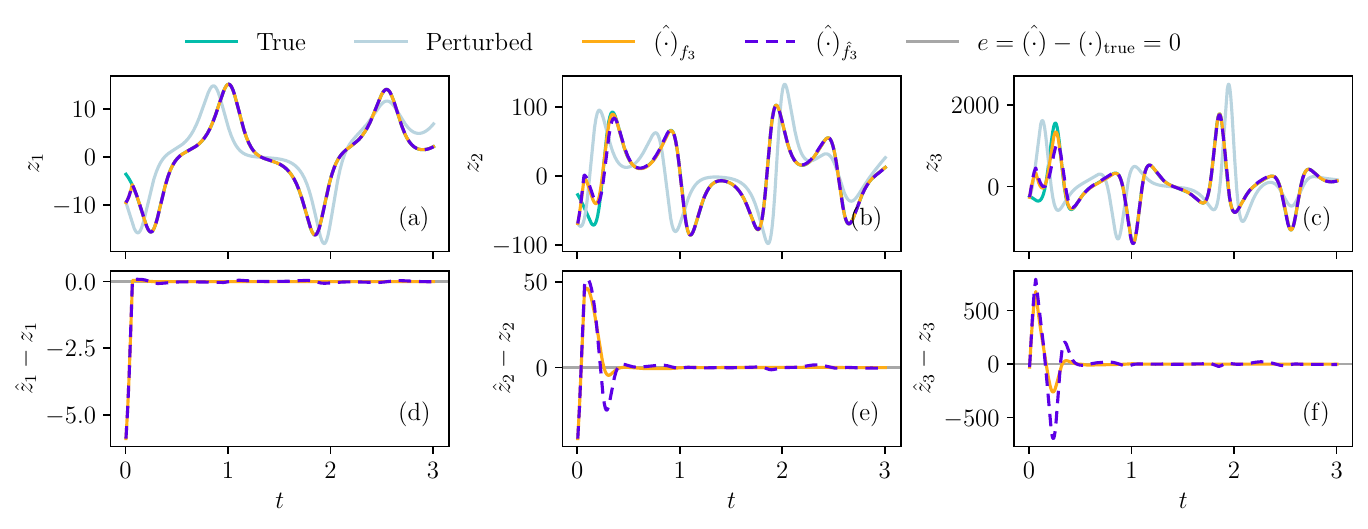}
    \caption{Sliding mode observer for the Lorenz system in the companion form with the true ODE \eqref{eq:lorenz_z3_dot}, $\hat{(\cdot)}_{f_3}$, and the learned neural ODE \eqref{eq:neural_ode}, $\hat{(\cdot)}_{\hat{f}_3}$ when the first state variable $x_1$ is observed.}\label{fig:lorenz_smo_h_x}
\end{figure}
\begin{figure}[tb]
    \includegraphics[width=\linewidth]{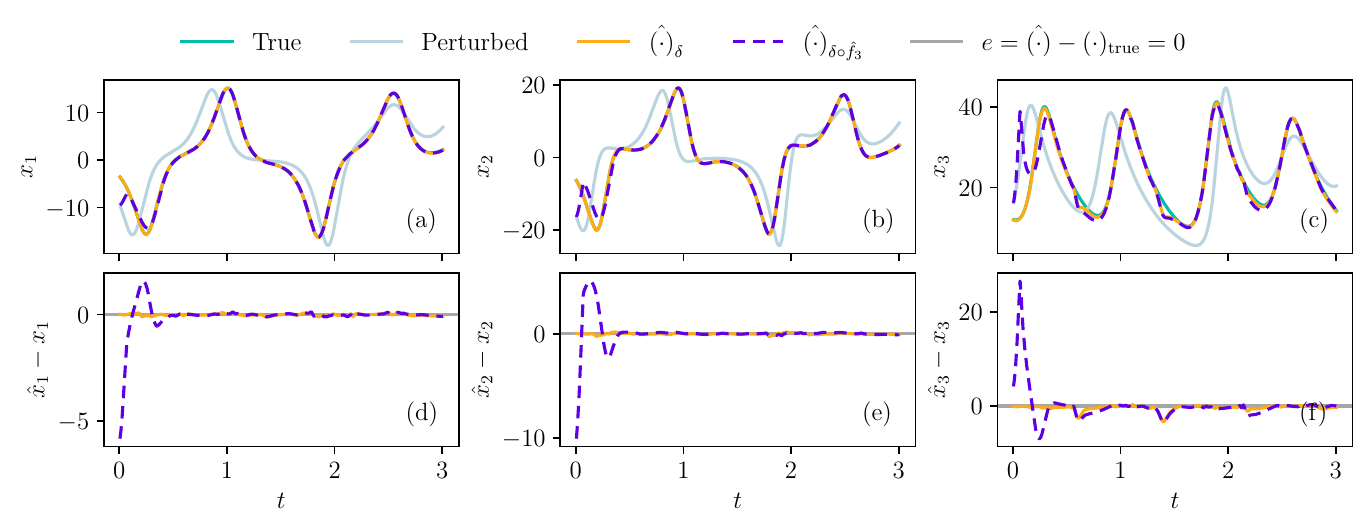}
    \caption{State estimation of the Lorenz system from $x_1$ observations by decoding \eqref{eq:decoder} the true $Z$, $\hat{(\cdot)}_{\delta}$, and decoding the estimation of the sliding mode observer $\hat{Z}$ with the learned neural ODE, $\hat{(\cdot)}_{\delta \circ \hat{f}_3}$.}\label{fig:lorenz_pred_x_nn}
\end{figure}

We perform the state estimation of the same trajectory in \S \ref{sec:lorenz_example} using the trained neural ODE in the companion form. We require $k_1 > |e_2|$ for sliding and the equivalent error dynamics are stable for any $k_2, k_3 > 0$. Additionally, we require $k_3 > |\Delta f_3|$ such that the error in the non-linear dynamics is dominated by the observer gain. Tuning these gains and the saturation threshold \eqref{eq:saturation}, we find $k_1 = 10^2, k_2 = 10^3, k_3 = 10^4, \epsilon = 10^{-1}$ provides a comparable performance. Figure \ref{fig:lorenz_smo_h_x} shows the successful state estimation for $Z$ using this sliding mode observer with: (a) the true ODE found analytically, $f_3$ \eqref{eq:lorenz_z3_dot}, and (b) the learned neural ODE $\hat{f}_3$. Figure \ref{fig:lorenz_pred_x_nn} compares decoded values of the estimated and true $Z$, that is $\delta(\hat{Z})$ and $\delta(Z)$, respectively, finding accurate alignment between the truth and estimate for the majority of the considered window. The above mentioned decoding error occurs in $x_3$ due to the loss of observability, which can be seen for example at $t = 1.5$ in Figure \ref{fig:lorenz_pred_x_nn} (c, f).
%\begin{figure}[tb]
%    \includegraphics[width=\linewidth]{graphics/lorenz_smo_h_y.pdf}
%    \caption{Sliding mode observer in the companion form with the true ODE, $\hat{(\cdot)}_{f_4}$, and the learned neural ODE, $\hat{(\cdot)}_{\hat{f}_4}$.}\label{fig:lorenz_smo_h_y}
%\end{figure}
%\begin{figure}[tb]
%    \includegraphics[width=\linewidth]{graphics/lorenz_pred_y_nn.pdf}
%   \caption{State estimation of the Lorenz system from $x$ observations by decoding the true $\vect{H}_4$, $\hat{(\cdot)}_{\delta}$, and decoding the $\vect{H}_4$ estimation of the sliding mode observer with the learned neural ODE, $\hat{(\cdot)}_{\delta \circ \hat{f}_4}$.}\label{fig:lorenz_pred_y_nn}
%\end{figure}
Repeating the same training procedure for $x_2$ observations, we find that adding the third time derivative as an input, i.e., $Z = (z_1, z_2, z_3, z_4)^\top = (x_2, \dot{x}_2, \ddot{x}_2, x_2^{(3)})^\top$, significantly improves the decoder error ($\sim 80\%$ in the validation error). This is in line with our previous analysis and numerical results that for $x_2$ observations, including the third Lie derivative gives distinct states. In this case, for the stability of the observer, we require $k_2 k_3 > k_1 k_4$ using the Hurwitz condition, in addition to $k_1 > |e_2|$ and $k_4 > |\Delta f_4|$. For example, the gains $k_1 = 10^2, k_2 = 10^4, k_3 = 10^4,  k_4 = 10^5$ satisfy these conditions and lead to the convergence of the state estimation error. %The state estimation for $\vect{H}_4$ using this sliding mode observer, and the decoded values following the estimation are shown in Figures \ref{fig:lorenz_smo_h_y} and \ref{fig:lorenz_pred_y_nn}, respectively. 
As for $x_3$ observations, due to the loss of observability, the decoder cannot find a valid mapping from the observations and their time derivatives to the full state aligning with our previous results. 
\subsection{Results for a nine dimensional model of a turbulent shear flow}\label{sec:results_mfe}
\begin{figure}[tb]
    \includegraphics[width=\linewidth]{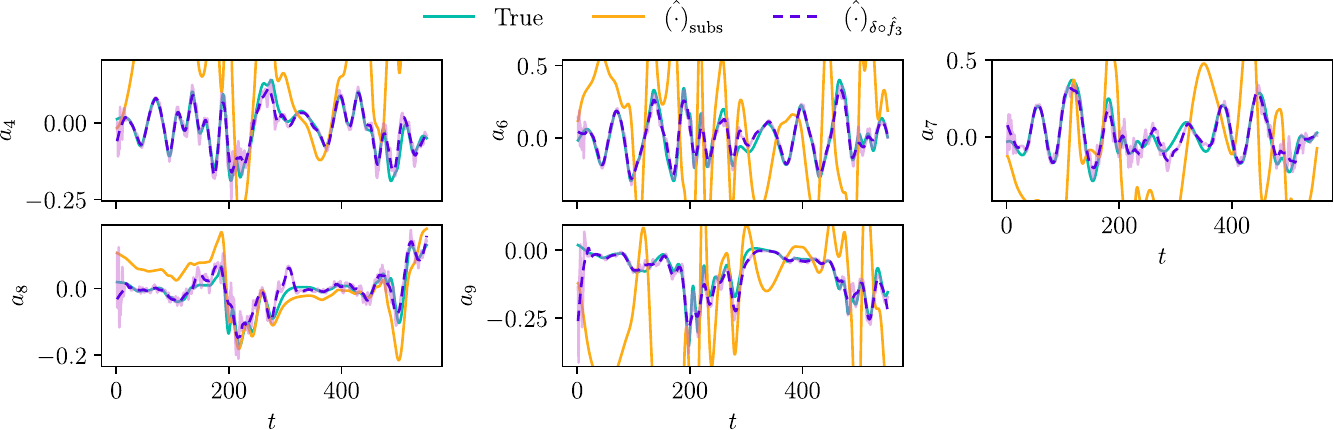}
    \caption{State estimation of unobserved five modes in the nine-mode system. In contrast to direct substitution $\hat{(\cdot)}_{\text{subs}}$, the proposed sliding mode observer based on neural approximations $\hat{(\cdot)}_{\text{smo}}$ achieves synchronisation. (Shaded area shows the decoder prediction before filtering).}\label{fig:mfe_pred_nn_obs_modes_1235}
\end{figure}
For the three-dimensional Lorenz system, we were able to find analytical expressions for the companion form dynamics and their transformation back to the original physical space. In this section, we apply the neural network approach to the nine-dimensional model of a turbulent shear flow discussed in \S\ref{sec:mfe}, for which we do not have such an expression. We generate data for training, validation and testing by perturbing the base initial condition $a_0 = \hat{\boldsymbol{e}}_1 + 0.7066 \hat{\boldsymbol{e}}_3 - 0.07606 \hat{\boldsymbol{e}}_4$ as $a_0' = a_0 + \epsilon \hat{\mathrm{\vect{e}}}_4$ where $\epsilon \sim\mathcal{U}(0,0.01)$ is a random variable drawn from a uniform distribution, similarly to \citep{racca_data-driven_2022}. We simulate 30 such trajectories for training and 8 for validation. We discard the first 200 time units ($\approxeq 3$ LTs) of each trajectory, such that they are sufficiently decorrelated from each other. After this operation, the remaining trajectories are 4,000 time units ($\approxeq 65$ LTs) long. We focus on a case, for which we could not find a feasible solution via the SDP approach. We consider the problem of observing four modes out of nine such that $Ca = (a_1, a_2, a_3, a_5)^\top$ with $p = 4$ and reconstructing the remaining five modes. Taking the first two time derivatives of the observations, we create a latent space with a dimension of 12, which is greater than the original space dimension of $n = 9$. (Similarly to the Lorenz data, the time derivatives are approximated using a fourth-order numerical differentiation scheme, and the inputs are standardised.) In this case, due to the increasing problem size, we construct larger networks with two layers of 256 neurons each using $\tanh$ activation. As discussed above, we perform the training with a batch size of one with Adam optimizer at a learning rate of $10^{-4}$ for 10 epochs, at which point the validation loss saturates. Following training, we design a sliding mode observer in the transformed space in companion form, where the gains corresponding to different observed modes are decoupled such that we have $K = (0.1I_4 \;\; 0.1I_4 \;\; 0.05I_4)^\top$. (The saturation threshold \eqref{eq:saturation} is set to $\epsilon = 0.01$.) With these gains, the observer rapidly converges to the true state. The decoded predictions for a random trajectory are shown in Figure \ref{fig:mfe_pred_nn_obs_modes_1235} in comparison to the direct substitution approach. The data-driven sliding mode observer significantly outperforms the direct substitution method, enabling the synchronisation of the unobserved variables. The decoding stage introduces noise, which may be due to the numerical differentiation in the training step or an indicator of loss of observability. For clarity in the plots, we smooth out the noisy predictions with a Savitsky-Golay filter.

\begin{figure}[tb]
    \includegraphics[width=\linewidth]{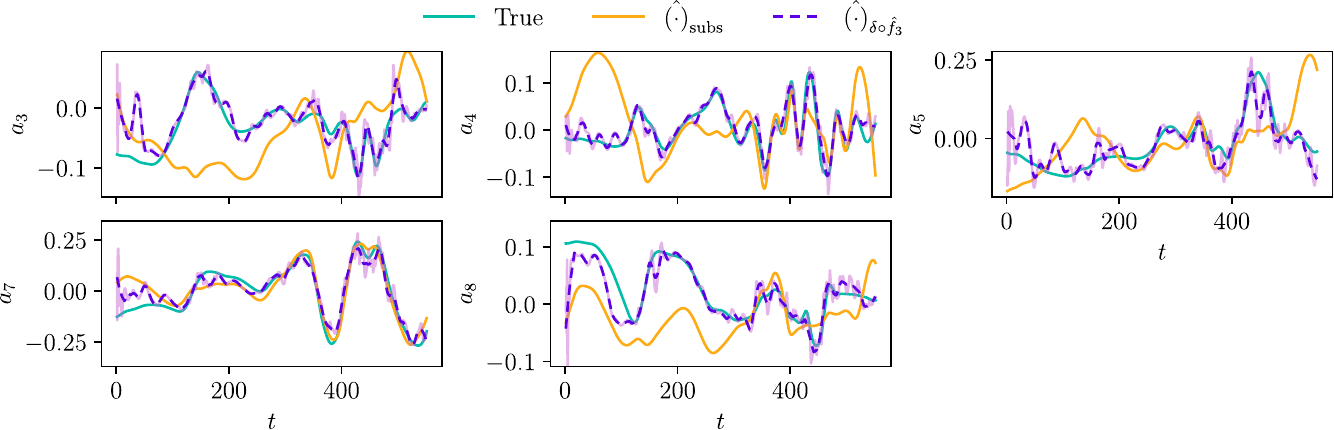}
    \caption{State estimation of the nine-mode system from probe measurements of the streamwise velocity. As this sensor setup is equivalent to observing the modes $a_1$, $a_2$, $a_6$, $a_9$, only the estimation of the remaining modes is shown. (See Figure \ref{fig:mfe_pred_nn_obs_modes_1235}).}\label{fig:mfe_pred_nn_obs_probes_pod}
\end{figure}
In experiments, we typically have measurements from probes rather than observing the modes themselves. Next, we consider such a case, where velocity measurements on a plane inside the domain are given. Given a $4 \times 4$ grid of $u$ measurements, i.e., streamwise velocity, on $x = 0$ plane, our goal is to reconstruct the nine mode coefficients. Although we have 16 measurements, the observation matrix associated with this setup has rank four. Therefore, after removing the mean, we apply a proper orthogonal decomposition (POD) on the measurements while retaining four POD modes corresponding to $99.9\%$ of the energy. These four retained POD modes and their time derivatives up to second order create a latent space with a dimension of 12. We perform the same training procedure as the previous case using the same architectures and hyperparameters. For this case, the gains $K = (0.8I_4 \;\; 0.8I_4 \;\; 0.4I_4)^\top$ lead to convergence in the transformed space. (The saturation threshold \eqref{eq:saturation} is set to $\epsilon = 0.1$.) In this case, the application of the direct substitution method is less straightforward. We employ the linear state transformation \eqref{eq:state_transform} introduced in \S\ref{sec:general_equiv_err_rom}. Taking measurements on a plane, i.e., evaluating the physical modes at the measurement locations and taking their sum, and the POD basis transformation are both linear operations. We obtain a linear observation matrix $C$ by multiplying them and accounting for the mean of the measurements. Upon this transformation, the direct substitution can be applied as before, where the first four state variables are substituted with the observed POD coefficients. Because the entries corresponding to the modes $a_1$, $a_2$, $a_6$, $a_9$ dominate the rank four observation matrix, both methods yield quick convergence to the truth in these modes. 
%\begin{equation*}
%    C = \begin{bmatrix}
%        4.45 & -1.66 \times 10^{-3} & 0 & 0 & 0 & -7.83 \times 10^{-3} & 0 & 0 & -8.26 \times 10^{-1} \\
%       -1.18 \times 10^{-2} & 3.61 \times 10^{-2} & 0 & 0 & 0 & -4.25 & 0 & 0 & 2.73 \times 10^{-2} \\
%        -6.35 \times 10^{-3} & -3.93 & 0 & 0 & 0 & -3.89 \times 10^{-2} & 0 & 0 & 3.17 \times 10^{-2} \\
%        7.73 \times 10^{-1} & 2.21 \times 10^{-2} & 0 & 0 & 0 & 2.01 \times 10^{-2} & 0 & 0 & 5.51
%    \end{bmatrix},
%\end{equation*}
\begin{figure}[tb]
    \centering
    \includegraphics[width=0.9\linewidth]{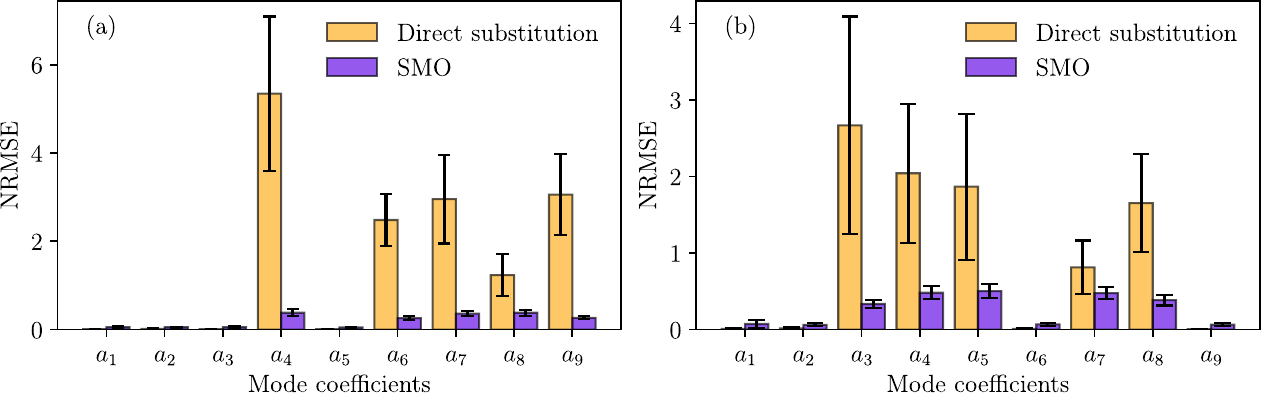}
    \caption{Estimation error statistics using the normalised root mean squared error (NRMSE): (a) Mode observations of $a_1$, $a_2$, $a_3$, $a_5$, and (b) probe observations.}\label{fig:error_comparison}
\end{figure}
This indicates that observing probes can become equivalent to observing certain modes. We show the state estimation for the remaining unobserved modes, $a_3$, $a_4$, $a_5$, $a_7$,   $a_8$ for a random trajectory in Figure \ref{fig:mfe_pred_nn_obs_probes_pod}. For these modes, in contrast to the direct substitution approach, the neural network based sliding mode observer enables synchronisation, albeit with some noise introduced at the decoding stage similarly to the previous case. For both observation configurations, the performances of the two approaches are quantified using the normalised root mean squared error (NRMSE) over 20 trajectories that are 4,000 time units long, the statistics of which are provided in Figure \ref{fig:error_comparison}. While direct substitutions results in zero error in the observed variables as expected, these two configurations exemplify where pure synchronisation fails. Assisted synchronisation utilising learned neural maps consistently achieves lower estimation errors in the unobserved variables across different trajectories and initial conditions. 
\section{Conclusion}\label{sec:conclusion}
The state estimation of turbulent flows remains challenging due to the underlying chaos. In this paper, we introduce a new control-theoretic perspective to systematically design observers for turbulent flows and demonstrate it on low-dimensional prototypical chaotic systems that qualitatively model turbulent behaviour. We first show that chaos synchronisation based on direct substitution of state variables can be regarded as a special type of sliding mode observer, which has enabled us to (i) analytically verify the asymptotic stability of the estimation error, and (ii) design feedback mechanisms for improved estimation performance. For this purpose, we propose two methods: (i) a semidefinite programming (SDP) approach based on sampling the data, and (ii) a neural network approach based on learning data-driven mappings. While the SDP approach is suitable for a large class of models in fluids with quadratic non-linearities, the neural network approach can be employed for any kind of non-linearity and does not require a physics-based model. The advantage of the SDP approach is that it is robust and can be directly employed for any linear observation matrix, whereas the neural network approach is affected by the loss of observability and requires retraining for any new observation matrix. The results show that either approach has the potential to significantly outperform the basic direct substitution method, while offering control-theoretic guarantees. Future research directions include employing higher-order and time-varying sliding mode observer designs to improve performance, investigating the source of the decoder noise in relation to the observability of higher-dimensional systems, and extension of the methods to higher-dimensional systems. This work opens opportunities in developing new data-driven methods for the state estimation of turbulent flows.\\
\\
\noindent\textbf{Funding} \hspace{0.05em}
This work was supported by the Elsie Widdowson Fellowship Award granted by Imperial College London; and was partially funded by UKRI under grant UKRI4171.
\vspace{0.5em}

\noindent\textbf{Acknowledgments} \hspace{0.05em}
The authors acknowledge computational resources and support provided by the Imperial College Research Computing Service (http://doi.org/10.14469/hpc/2232).
\vspace{0.5em}

\noindent\textbf{Declaration of use of generative AI and AI-assisted technologies} \hspace{0.05em}
The authors acknowledge the use of GPT-based tools to assist with debugging and streamlining the code. After using these tools, the authors reviewed and edited the code as needed and take full responsibility for the content of the published article.
\vspace{0.5em}

\noindent\textbf{Data availability} \hspace{0.05em}
Data will be made available publicly upon publication.
\appendix
\section{Nine-dimensional model of a turbulent shear flow}\label{sec:appendix}
The governing equations for the nine-dimensional model of a turbulent shear flow are given by \citep{moehlis_low-dimensional_2004}
\begin{equation}\label{eq:mfe}
    \begin{split}
\dot a_1 &= \frac{\beta^2}{Re} - \frac{\beta^2}{Re} a_1
- \sqrt{\frac{3}{2}}\,\frac{\beta\gamma}{\kappa_{\alpha\beta\gamma}}\, a_6 a_8
+ \sqrt{\frac{3}{2}}\,\frac{\beta\gamma}{\kappa_{\beta\gamma}}\, a_2 a_3,
\\[0.75em]
\dot a_2 &= -\frac{\tfrac{4}{3}\beta^2+\gamma^2}{Re}\, a_2
+ \frac{5\sqrt{2}}{3\sqrt{3}}\,\frac{\gamma^2}{\kappa_{\alpha\gamma}}\, a_4 a_6
- \frac{\gamma^2}{\sqrt{6}\,\kappa_{\alpha\gamma}}\, a_5 a_7
 - \frac{\alpha\beta\gamma}{\sqrt{6}\,\kappa_{\alpha\gamma}\kappa_{\alpha\beta\gamma}}\, a_5 a_8
- \sqrt{\frac{3}{2}}\,\frac{\beta\gamma}{\kappa_{\beta\gamma}}\, a_1 a_3
- \sqrt{\frac{3}{2}}\,\frac{\beta\gamma}{\kappa_{\beta\gamma}}\, a_3 a_9,
\\[0.75em]
\dot a_3 &= -\frac{\beta^2+\gamma^2}{Re}\, a_3
+ \frac{2}{\sqrt{6}}\,\frac{\alpha\beta\gamma}{\kappa_{\alpha\gamma}\kappa_{\beta\gamma}}\,(a_4 a_7 + a_5 a_6)
+ \frac{\beta^2(3\alpha^2+\gamma^2)-3\gamma^2(\alpha^2+\gamma^2)}
{\sqrt{6}\,\kappa_{\alpha\gamma}\kappa_{\beta\gamma}\kappa_{\alpha\beta\gamma}}\, a_4 a_8,
\\[0.75em]
\dot a_4 &= -\frac{3\alpha^2+4\beta^2}{3Re}\, a_4
- \frac{\alpha}{\sqrt{6}}\,(a_1 a_5 + a_5 a_9)
- \frac{10\alpha^2}{3\sqrt{6}\,\kappa_{\alpha\gamma}}\, a_2 a_6
- \sqrt{\frac{3}{2}}\,\frac{\alpha\beta\gamma}{\kappa_{\alpha\gamma}\kappa_{\beta\gamma}}\, a_3 a_7
- \sqrt{\frac{3}{2}}\,\frac{\alpha^2\beta^2}{\kappa_{\alpha\gamma}\kappa_{\beta\gamma}\kappa_{\alpha\beta\gamma}}\, a_3 a_8,
\\[0.75em]
\dot a_5 &= -\frac{\alpha^2+\beta^2}{Re}\, a_5
+ \frac{\alpha}{\sqrt{6}}\,(a_1 a_4 + a_4 a_9)
+ \frac{\alpha^2}{\sqrt{6}\,\kappa_{\alpha\gamma}}\, a_2 a_7
- \frac{\alpha\beta\gamma}{\sqrt{6}\,\kappa_{\alpha\gamma}\kappa_{\alpha\beta\gamma}}\, a_2 a_8
+ \frac{2}{\sqrt{6}}\,\frac{\alpha\beta\gamma}{\kappa_{\alpha\gamma}\kappa_{\beta\gamma}}\, a_3 a_6,
\\[0.75em]
\dot a_6 &= -\frac{3\alpha^2+4\beta^2+3\gamma^2}{3Re}\, a_6
+ \frac{\alpha}{\sqrt{6}}\,(a_1 a_7 + a_7 a_9)
+ \sqrt{\frac{3}{2}}\,\frac{\beta\gamma}{\kappa_{\alpha\beta\gamma}}\, a_1 a_8
 + \frac{10(\alpha^2-\gamma^2)}{3\sqrt{6}\,\kappa_{\alpha\gamma}}\, a_2 a_4
\\
&\quad - 2\sqrt{\frac{2}{3}}\,\frac{\alpha\beta\gamma}{\kappa_{\alpha\gamma}\kappa_{\beta\gamma}}\, a_3 a_5
+ \sqrt{\frac{3}{2}}\,\frac{\beta\gamma}{\kappa_{\alpha\beta\gamma}}\, a_8 a_9,
\\[0.75em]
\dot a_7 &= -\frac{\alpha^2+\beta^2+\gamma^2}{Re}\, a_7
- \frac{\alpha}{\sqrt{6}}\,(a_1 a_6 + a_6 a_9)
 + \frac{\gamma^2-\alpha^2}{\sqrt{6}\,\kappa_{\alpha\gamma}}\, a_2 a_5  + \frac{\alpha\beta\gamma}{\sqrt{6}\,\kappa_{\alpha\gamma}\kappa_{\beta\gamma}}\, a_3 a_4,
\\[0.75em]
\dot a_8 &= -\frac{\alpha^2+\beta^2+\gamma^2}{Re}\, a_8
+ \frac{2\alpha\beta\gamma}{\sqrt{6}\,\kappa_{\alpha\gamma}\kappa_{\alpha\beta\gamma}}\, a_2 a_5 + \frac{\gamma^2(3\alpha^2-\beta^2+3\gamma^2)}
{\sqrt{6}\,\kappa_{\alpha\gamma}\kappa_{\beta\gamma}\kappa_{\alpha\beta\gamma}}\, a_3 a_4,
\\[0.75em]
\dot a_9 &= -\frac{9\beta^2}{Re}\, a_9
+ \sqrt{\frac{3}{2}}\,\frac{\beta\gamma}{\kappa_{\beta\gamma}}\, a_2 a_3
- \sqrt{\frac{3}{2}}\,\frac{\beta\gamma}{\kappa_{\alpha\beta\gamma}}\, a_6 a_8.
    \end{split}
\end{equation}
where $\alpha = 2 \pi / L_x$, $\beta = \pi / 2$, $\gamma = 2 \pi / L_z$ are constants that depend on the dimensions of the domain described by $L_x$ and $L_z$. The auxiliary wavenumber combinations are defined by
\begin{equation*}
\kappa_{\alpha\gamma} = \sqrt{\alpha^2 + \gamma^2}, \;
\kappa_{\beta\gamma} = \sqrt{\beta^2 + \gamma^2}, \;
\kappa_{\alpha\beta\gamma} = \sqrt{\alpha^2 + \beta^2 + \gamma^2}.
\end{equation*}
\bibliographystyle{plainnat}
\bibliography{bibliography}

@article{cross_differential_2010,
	title = {Differential embedding of the {Lorenz} attractor},
	volume = {81},
	copyright = {http://link.aps.org/licenses/aps-default-license},
	issn = {1539-3755, 1550-2376},
	url = {https://link.aps.org/doi/10.1103/PhysRevE.81.066220},
	doi = {10.1103/PhysRevE.81.066220},
	language = {en},
	number = {6},
	urldate = {2026-05-18},
	journal = {Physical Review E},
	author = {Cross, Daniel J. and Gilmore, R.},
	month = jun,
	year = {2010},
	pages = {066220},
}

@article{moehlis_low-dimensional_2004,
	title = {A low-dimensional model for turbulent shear flows},
	volume = {6},
	issn = {1367-2630},
	url = {https://iopscience.iop.org/article/10.1088/1367-2630/6/1/056},
	doi = {10.1088/1367-2630/6/1/056},
	urldate = {2026-05-18},
	journal = {New Journal of Physics},
	author = {Moehlis, Jeff and Faisst, Holger and Eckhardt, Bruno},
	month = may,
	year = {2004},
	pages = {56--56},
}

@article{le_dimet_variational_1986,
	title = {Variational algorithms for analysis and assimilation of meteorological observations: theoretical aspects},
	volume = {38},
	issn = {1600-0870},
	shorttitle = {Variational algorithms for analysis and assimilation of meteorological observations},
	url = {https://tellusjournal.org/articles/10.3402/tellusa.v38i2.11706},
	doi = {10.3402/tellusa.v38i2.11706},
	number = {2},
	urldate = {2026-05-18},
	journal = {Tellus A: Dynamic Meteorology and Oceanography},
	author = {Le Dimet, François-Xavier and Talagrand, Olivier},
	month = jan,
	year = {1986},
	pages = {97},
}

@article{wang_state_2021,
	title = {State estimation in turbulent channel flow from limited observations},
	volume = {917},
	issn = {0022-1120, 1469-7645},
	url = {https://www.cambridge.org/core/product/identifier/S0022112021002688/type/journal_article},
	doi = {10.1017/jfm.2021.268},
	language = {en},
	urldate = {2026-05-18},
	journal = {Journal of Fluid Mechanics},
	author = {Wang, Mengze and Zaki, Tamer A.},
	month = jun,
	year = {2021},
	pages = {A9},
}

@article{suzuki_estimation_2017,
	title = {Estimation of turbulent channel flow at based on the wall measurement using a simple sequential approach},
	volume = {830},
	copyright = {https://www.cambridge.org/core/terms},
	issn = {0022-1120, 1469-7645},
	url = {https://www.cambridge.org/core/product/identifier/S0022112017005808/type/journal_article},
	doi = {10.1017/jfm.2017.580},
	language = {en},
	urldate = {2026-05-18},
	journal = {Journal of Fluid Mechanics},
	author = {Suzuki, Takao and Hasegawa, Yosuke},
	month = nov,
	year = {2017},
	pages = {760--796},
}

@article{symon_data_2017,
	title = {Data assimilation of mean velocity from {2D} {PIV} measurements of flow over an idealized airfoil},
	volume = {58},
	issn = {0723-4864, 1432-1114},
	url = {http://link.springer.com/10.1007/s00348-017-2336-8},
	doi = {10.1007/s00348-017-2336-8},
	language = {en},
	number = {5},
	urldate = {2026-05-18},
	journal = {Experiments in Fluids},
	author = {Symon, Sean and Dovetta, Nicolas and McKeon, Beverley J. and Sipp, Denis and Schmid, Peter J.},
	month = may,
	year = {2017},
	pages = {61},
}

@article{kelshaw_physics-constrained_2024,
	title = {Physics-constrained convolutional neural networks for inverse problems in spatiotemporal partial differential equations},
	volume = {5},
	issn = {2632-6736},
	url = {https://www.cambridge.org/core/product/identifier/S2632673624000467/type/journal_article},
	doi = {10.1017/dce.2024.46},
	language = {en},
	urldate = {2026-05-18},
	journal = {Data-Centric Engineering},
	author = {Kelshaw, Daniel and Magri, Luca},
	year = {2024},
	pages = {e43},
}

@article{mo_reconstructing_2025,
	title = {Reconstructing unsteady flows from sparse, noisy measurements with a physics-constrained convolutional neural network},
	volume = {10},
	issn = {2469-990X},
	url = {https://link.aps.org/doi/10.1103/PhysRevFluids.10.034901},
	doi = {10.1103/PhysRevFluids.10.034901},
	language = {en},
	number = {3},
	urldate = {2026-05-18},
	journal = {Physical Review Fluids},
	author = {Mo, Yaxin and Magri, Luca},
	month = mar,
	year = {2025},
	pages = {034901},
}

@article{gao_super-resolution_2021,
	title = {Super-resolution and denoising of fluid flow using physics-informed convolutional neural networks without high-resolution labels},
	volume = {33},
	issn = {1070-6631, 1089-7666},
	url = {https://pubs.aip.org/aip/pof/article/1076765},
	doi = {10.1063/5.0054312},
	language = {en},
	number = {7},
	urldate = {2026-05-18},
	journal = {Physics of Fluids},
	author = {Gao, Han and Sun, Luning and Wang, Jian-Xun},
	month = jul,
	year = {2021},
	pages = {073603},
}

@article{fukami_super-resolution_2019,
	title = {Super-resolution reconstruction of turbulent flows with machine learning},
	volume = {870},
	copyright = {https://www.cambridge.org/core/terms},
	issn = {0022-1120, 1469-7645},
	url = {https://www.cambridge.org/core/product/identifier/S0022112019002386/type/journal_article},
	doi = {10.1017/jfm.2019.238},
	language = {en},
	urldate = {2026-05-18},
	journal = {Journal of Fluid Mechanics},
	author = {Fukami, Kai and Fukagata, Koji and Taira, Kunihiko},
	month = jul,
	year = {2019},
	pages = {106--120},
}

@article{erichson_shallow_2020,
	title = {Shallow neural networks for fluid flow reconstruction with limited sensors},
	volume = {476},
	issn = {1364-5021, 1471-2946},
	url = {http://royalsocietypublishing.org/rspa/article/80735},
	doi = {10.1098/rspa.2020.0097},
	language = {en},
	number = {2238},
	urldate = {2026-05-18},
	journal = {Proceedings of the Royal Society A: Mathematical, Physical and Engineering Sciences},
	author = {Erichson, N. Benjamin and Mathelin, Lionel and Yao, Zhewei and Brunton, Steven L. and Mahoney, Michael W. and Kutz, J. Nathan},
	month = jun,
	year = {2020},
	pages = {20200097},
}

@article{nijmeijer_observer_1997,
	title = {An observer looks at synchronization},
	volume = {44},
	issn = {1558-1268},
	url = {https://ieeexplore.ieee.org/document/633877},
	doi = {10.1109/81.633877},
	number = {10},
	urldate = {2026-03-05},
	journal = {IEEE Transactions on Circuits and Systems I: Fundamental Theory and Applications},
	author = {Nijmeijer, H. and Mareels, I.M.Y.},
	month = oct,
	year = {1997},
	pages = {882--890},
}

@inproceedings{drakunov_sliding_1995,
	title = {Sliding mode observers. {Tutorial}},
	volume = {4},
	issn = {0191-2216},
	url = {https://ieeexplore.ieee.org/document/479009},
	doi = {10.1109/CDC.1995.479009},
	urldate = {2026-03-02},
	booktitle = {Proceedings of 1995 34th {IEEE} {Conference} on {Decision} and {Control}},
	author = {Drakunov, S. and Utkin, V.},
	month = dec,
	year = {1995},
	note = {ISSN: 0191-2216},
	pages = {3376--3378 vol.4},
}

@article{wang_variational_2025,
	title = {Variational data assimilation in wall turbulence: from outer observations to wall stress and pressure},
	volume = {1008},
	issn = {0022-1120, 1469-7645},
	shorttitle = {Variational data assimilation in wall turbulence},
	url = {https://www.cambridge.org/core/product/identifier/S0022112025001326/type/journal_article},
	doi = {10.1017/jfm.2025.132},
	language = {en},
	urldate = {2026-02-18},
	journal = {Journal of Fluid Mechanics},
	author = {Wang, Mengze and Zaki, Tamer A.},
	month = apr,
	year = {2025},
	pages = {A26},
}

@article{li_small-scale_2020,
	title = {Small-scale reconstruction in three-dimensional {Kolmogorov} flows using four-dimensional variational data assimilation},
	volume = {885},
	copyright = {https://www.cambridge.org/core/terms},
	issn = {0022-1120, 1469-7645},
	url = {https://www.cambridge.org/core/product/identifier/S0022112019009601/type/journal_article},
	doi = {10.1017/jfm.2019.960},
	language = {en},
	urldate = {2026-02-18},
	journal = {Journal of Fluid Mechanics},
	author = {Li, Yi and Zhang, Jianlei and Dong, Gang and Abdullah, Naseer S.},
	month = feb,
	year = {2020},
	pages = {A9},
}

@article{lalescu_synchronization_2013,
	title = {Synchronization of {Chaos} in {Fully} {Developed} {Turbulence}},
	volume = {110},
	copyright = {http://link.aps.org/licenses/aps-default-license},
	issn = {0031-9007, 1079-7114},
	url = {https://link.aps.org/doi/10.1103/PhysRevLett.110.084102},
	doi = {10.1103/PhysRevLett.110.084102},
	language = {en},
	number = {8},
	urldate = {2026-02-18},
	journal = {Physical Review Letters},
	author = {Lalescu, Cristian C. and Meneveau, Charles and Eyink, Gregory L.},
	month = feb,
	year = {2013},
	pages = {084102},
}

@article{yoshida_regeneration_2005,
	title = {Regeneration of {Small} {Eddies} by {Data} {Assimilation} in {Turbulence}},
	volume = {94},
	copyright = {http://link.aps.org/licenses/aps-default-license},
	issn = {0031-9007, 1079-7114},
	url = {https://link.aps.org/doi/10.1103/PhysRevLett.94.014501},
	doi = {10.1103/PhysRevLett.94.014501},
	language = {en},
	number = {1},
	urldate = {2026-02-18},
	journal = {Physical Review Letters},
	author = {Yoshida, Kyo and Yamaguchi, Junzo and Kaneda, Yukio},
	month = jan,
	year = {2005},
	pages = {014501},
}

@article{wang_synchronization_2022,
	title = {Synchronization of turbulence in channel flow},
	volume = {943},
	issn = {0022-1120, 1469-7645},
	url = {https://www.cambridge.org/core/product/identifier/S0022112022003974/type/journal_article},
	doi = {10.1017/jfm.2022.397},
	language = {en},
	urldate = {2026-02-18},
	journal = {Journal of Fluid Mechanics},
	author = {Wang, Mengze and Zaki, Tamer A.},
	month = jul,
	year = {2022},
	pages = {A4},
}

@article{pecora_synchronization_1990,
	title = {Synchronization in chaotic systems},
	volume = {64},
	url = {https://link.aps.org/doi/10.1103/PhysRevLett.64.821},
	doi = {10.1103/PhysRevLett.64.821},
	number = {8},
	urldate = {2026-02-18},
	journal = {Physical Review Letters},
	publisher = {American Physical Society},
	author = {Pecora, Louis M. and Carroll, Thomas L.},
	month = feb,
	year = {1990},
	pages = {821--824},
}

@inproceedings{slotine_sliding_1986,
	title = {On {Sliding} {Observers} for {Nonlinear} {Systems}},
	url = {https://ieeexplore.ieee.org/abstract/document/4789217},
	doi = {10.23919/ACC.1986.4789217},
	urldate = {2026-01-14},
	booktitle = {1986 {American} {Control} {Conference}},
	author = {Slotine, J.-J. E. and Hedrick, J. K. and Misawa, E. A.},
	month = jun,
	year = {1986},
	pages = {1794--1800},
}

@article{page_super-resolution_2025,
	title = {Super-resolution of turbulence with dynamics in the loss},
	volume = {1002},
	copyright = {http://creativecommons.org/licenses/by/4.0},
	issn = {0022-1120, 1469-7645},
	url = {https://www.cambridge.org/core/product/identifier/S0022112024012023/type/journal_article},
	doi = {10.1017/jfm.2024.1202},
	language = {en},
	urldate = {2025-07-12},
	journal = {Journal of Fluid Mechanics},
	publisher = {Cambridge University Press (CUP)},
	author = {Page, Jacob},
	month = jan,
	year = {2025},
}

@article{luenberger_introduction_1971,
	title = {An introduction to observers},
	volume = {16},
	copyright = {https://ieeexplore.ieee.org/Xplorehelp/downloads/license-information/IEEE.html},
	issn = {0018-9286},
	url = {http://ieeexplore.ieee.org/document/1099826/},
	doi = {10.1109/TAC.1971.1099826},
	language = {en},
	number = {6},
	urldate = {2025-05-07},
	journal = {IEEE Transactions on Automatic Control},
	author = {Luenberger, D.},
	month = dec,
	year = {1971},
	pages = {596--602},
}

@article{rowley_model_2017,
	title = {Model {Reduction} for {Flow} {Analysis} and {Control}},
	volume = {49},
	issn = {0066-4189, 1545-4479},
	url = {https://www.annualreviews.org/doi/10.1146/annurev-fluid-010816-060042},
	doi = {10.1146/annurev-fluid-010816-060042},
	language = {en},
	number = {1},
	urldate = {2025-05-29},
	journal = {Annual Review of Fluid Mechanics},
	author = {Rowley, Clarence W. and Dawson, Scott T.M.},
	month = jan,
	year = {2017},
	pages = {387--417},
}

@book{evensen_data_2009,
	address = {Berlin, Heidelberg},
	title = {Data {Assimilation}: {The} {Ensemble} {Kalman} {Filter}},
	copyright = {https://www.springernature.com/gp/researchers/text-and-data-mining},
	isbn = {978-3-642-03710-8 978-3-642-03711-5},
	shorttitle = {Data {Assimilation}},
	url = {https://link.springer.com/10.1007/978-3-642-03711-5},
	doi = {10.1007/978-3-642-03711-5},
	language = {en},
	urldate = {2025-02-15},
	publisher = {Springer Berlin Heidelberg},
	author = {Evensen, Geir},
	year = {2009},
}

@article{fossella_multiscale_2026,
	title = {Multiscale data assimilation in turbulent models},
	volume = {113},
	url = {https://link.aps.org/doi/10.1103/c1wn-n1j6},
	doi = {10.1103/c1wn-n1j6},
	number = {2},
	urldate = {2026-08-04},
	journal = {Physical Review E},
	publisher = {American Physical Society},
	author = {Fossella, Francesco and Biferale, Luca and Carrassi, Alberto and Cencini, Massimo and Gupta, Vikrant},
	month = feb,
	year = {2026},
	pages = {024208},
}

@article{hermann_nonlinear_1977,
	title = {Nonlinear controllability and observability},
	volume = {22},
	issn = {1558-2523},
	url = {https://ieeexplore.ieee.org/document/1101601},
	doi = {10.1109/TAC.1977.1101601},
	number = {5},
	urldate = {2026-08-04},
	journal = {IEEE Transactions on Automatic Control},
	author = {Hermann, R. and Krener, A.},
	month = oct,
	year = {1977},
	pages = {728--740},
}

@article{so_observing_1994,
	title = {Observing chaos: {Deducing} and tracking the state of a chaotic system from limited observation},
	volume = {49},
	shorttitle = {Observing chaos},
	url = {https://link.aps.org/doi/10.1103/PhysRevE.49.2650},
	doi = {10.1103/PhysRevE.49.2650},
	number = {4},
	urldate = {2026-08-04},
	journal = {Physical Review E},
	publisher = {American Physical Society},
	author = {So, Paul and Ott, Edward and Dayawansa, W. P.},
	month = apr,
	year = {1994},
	pages = {2650--2660},
}

@article{dirix_synthesis_2022,
	title = {Synthesis of patient-specific multipoint {4D} flow {MRI} data of turbulent aortic flow downstream of stenotic valves},
	volume = {12},
	copyright = {2022 The Author(s)},
	issn = {2045-2322},
	url = {https://www.nature.com/articles/s41598-022-20121-x},
	doi = {10.1038/s41598-022-20121-x},
	language = {en},
	number = {1},
	urldate = {2026-09-18},
	journal = {Scientific Reports},
	publisher = {Nature Publishing Group},
	author = {Dirix, Pietro and Buoso, Stefano and Peper, Eva S. and Kozerke, Sebastian},
	month = sep,
	year = {2022},
	pages = {16004},
}

@article{kato_data_2015,
	title = {A data assimilation methodology for reconstructing turbulent flows around aircraft},
	volume = {283},
	issn = {0021-9991},
	url = {https://www.sciencedirect.com/science/article/pii/S0021999114008213},
	doi = {10.1016/j.jcp.2014.12.013},
	urldate = {2026-09-18},
	journal = {Journal of Computational Physics},
	author = {Kato, Hiroshi and Yoshizawa, Akira and Ueno, Genta and Obayashi, Shigeru},
	month = feb,
	year = {2015},
	pages = {559--581},
}

@article{clark_di_leoni_synchronization_2020,
	title = {Synchronization to {Big} {Data}: {Nudging} the {Navier}-{Stokes} {Equations} for {Data} {Assimilation} of {Turbulent} {Flows}},
	volume = {10},
	shorttitle = {Synchronization to {Big} {Data}},
	url = {https://link.aps.org/doi/10.1103/PhysRevX.10.011023},
	doi = {10.1103/PhysRevX.10.011023},
	number = {1},
	urldate = {2026-09-18},
	journal = {Physical Review X},
	publisher = {American Physical Society},
	author = {Clark Di Leoni, Patricio and Mazzino, Andrea and Biferale, Luca},
	month = feb,
	year = {2020},
	pages = {011023},
}

@article{carrassi_data_2018,
	title = {Data assimilation in the geosciences: {An} overview of methods, issues, and perspectives},
	volume = {9},
	issn = {1757-7799},
	shorttitle = {Data assimilation in the geosciences},
	url = {https://wires.onlinelibrary.wiley.com/doi/10.1002/wcc.535},
	doi = {10.1002/wcc.535},
	language = {en},
	number = {5},
	urldate = {2026-09-18},
	journal = {Wiley Interdisciplinary Reviews: Climate Change},
	publisher = {John Wiley \& Sons, Ltd},
	author = {Carrassi, Alberto and Bocquet, Marc and Bertino, Laurent and Evensen, Geir},
	month = sep,
	year = {2018},
	pages = {e535},
}

@article{chevalier_state_2006,
	title = {State estimation in wall-bounded flow systems. {Part} 2. {Turbulent} flows},
	volume = {552},
	issn = {1469-7645, 0022-1120},
	url = {https://www.cambridge.org/core/journals/journal-of-fluid-mechanics/article/abs/state-estimation-in-wallbounded-flow-systems-part-2-turbulent-flows/CCA37DDF65BE5A1948835BA1455C4CC3},
	doi = {10.1017/S0022112005008578},
	language = {en},
	urldate = {2026-09-18},
	journal = {Journal of Fluid Mechanics},
	author = {Chevalier, Mattias and Hœpffner, Jérôme and Bewley, Thomas R. and Henningson, Dan S.},
	month = apr,
	year = {2006},
	pages = {167--187},
}

@inproceedings{lofberg_yalmip_2004,
	title = {{YALMIP} : a toolbox for modeling and optimization in {MATLAB}},
	shorttitle = {{YALMIP}},
	url = {https://ieeexplore.ieee.org/document/1393890},
	doi = {10.1109/CACSD.2004.1393890},
	urldate = {2026-09-24},
	booktitle = {2004 {IEEE} {International} {Conference} on {Robotics} and {Automation}},
	author = {Lofberg, J.},
	month = sep,
	year = {2004},
	pages = {284--289},
}

@book{mosek_aps_mosek_2026,
	address = {Copenhagen, Denmark},
	edition = {11.2.4},
	title = {{MOSEK} {Optimization} {Suite}},
	url = {https://docs.mosek.com/11.2/toolbox/index.html},
	publisher = {MOSEK ApS},
	author = {{MOSEK ApS}},
	year = {2026},
}

@article{srinivasan_predictions_2019,
	title = {Predictions of turbulent shear flows using deep neural networks},
	volume = {4},
	doi = {10.1103/PhysRevFluids.4.054603},
	number = {5},
	journal = {Physical Review Fluids},
	author = {Srinivasan, P. A.},
	year = {2019},
}

@article{skufca_edge_2006,
	title = {Edge of {Chaos} in a {Parallel} {Shear} {Flow}},
	volume = {96},
	url = {https://link.aps.org/doi/10.1103/PhysRevLett.96.174101},
	doi = {10.1103/PhysRevLett.96.174101},
	number = {17},
	urldate = {2026-09-24},
	journal = {Physical Review Letters},
	publisher = {American Physical Society},
	author = {Skufca, Joseph D. and Yorke, James A. and Eckhardt, Bruno},
	month = may,
	year = {2006},
	pages = {174101},
}

@article{racca_data-driven_2022,
	title = {Data-driven prediction and control of extreme events in a chaotic flow},
	volume = {7},
	url = {https://link.aps.org/doi/10.1103/PhysRevFluids.7.104402},
	doi = {10.1103/PhysRevFluids.7.104402},
	number = {10},
	urldate = {2026-09-24},
	journal = {Physical Review Fluids},
	publisher = {American Physical Society},
	author = {Racca, Alberto and Magri, Luca},
	month = oct,
	year = {2022},
	pages = {104402},
}

@inproceedings{chen_neural_2018,
	title = {Neural {Ordinary} {Differential} {Equations}},
	volume = {31},
	url = {https://proceedings.neurips.cc/paper_files/paper/2018/hash/69386f6bb1dfed68692a24c8686939b9-Abstract.html},
	urldate = {2026-11-06},
	booktitle = {Advances in {Neural} {Information} {Processing} {Systems}},
	publisher = {Curran Associates, Inc.},
	author = {Chen, Ricky T. Q. and Rubanova, Yulia and Bettencourt, Jesse and Duvenaud, David K},
	year = {2018},
}

@inproceedings{wynn_observer_2011,
	title = {Observer design for systems with an energy preserving nonlinearity, with application to fluid flow},
	issn = {0743-1546},
	url = {https://ieeexplore.ieee.org/abstract/document/6160445},
	doi = {10.1109/CDC.2011.6160445},
	urldate = {2026-11-03},
	booktitle = {2011 50th {IEEE} {Conference} on {Decision} and {Control} and {European} {Control} {Conference}},
	author = {Wynn, Andrew and Goulart, Paul},
	month = dec,
	year = {2011},
	note = {ISSN: 0743-1546},
	pages = {7524--7529},
}

@book{boyd_linear_1994,
	title = {Linear {Matrix} {Inequalities} in {System} and {Control} {Theory}},
	isbn = {978-0-89871-485-2 978-1-61197-077-7},
	url = {https://epubs.siam.org/doi/book/10.1137/1.9781611970777},
	doi = {10.1137/1.9781611970777},
	language = {en},
	urldate = {2026-11-03},
	publisher = {Society for Industrial and Applied Mathematics},
	author = {Boyd, Stephen and El Ghaoui, Laurent and Feron, Eric and Balakrishnan, Venkataramanan},
	month = jan,
	year = {1994},
}

@article{azouani_continuous_2014,
	title = {Continuous {Data} {Assimilation} {Using} {General} {Interpolant} {Observables}},
	volume = {24},
	copyright = {http://www.springer.com/tdm},
	issn = {0938-8974, 1432-1467},
	url = {http://link.springer.com/10.1007/s00332-013-9189-y},
	doi = {10.1007/s00332-013-9189-y},
	language = {en},
	number = {2},
	urldate = {2026-11-10},
	journal = {Journal of Nonlinear Science},
	author = {Azouani, Abderrahim and Olson, Eric and Titi, Edriss S.},
	month = apr,
	year = {2014},
	pages = {277--304},
}

@article{zeitz_observability_1984,
	title = {Observability canonical (phase-variable) form for non-linear time-variable systems},
	volume = {15},
	issn = {0020-7721, 1464-5319},
	url = {http://www.tandfonline.com/doi/abs/10.1080/00207728408926614},
	doi = {10.1080/00207728408926614},
	language = {en},
	number = {9},
	urldate = {2026-11-11},
	journal = {International Journal of Systems Science},
	author = {Zeitz, M.},
	month = sep,
	year = {1984},
	pages = {949--958},
}

@article{spurgeon_sliding_2008,
	title = {Sliding mode observers: a survey},
	volume = {39},
	issn = {0020-7721, 1464-5319},
	shorttitle = {Sliding mode observers},
	url = {http://www.tandfonline.com/doi/abs/10.1080/00207720701847638},
	doi = {10.1080/00207720701847638},
	language = {en},
	number = {8},
	urldate = {2026-11-16},
	journal = {International Journal of Systems Science},
	author = {Spurgeon, Sarah K.},
	month = aug,
	year = {2008},
	pages = {751--764},
}
\end{document}